\documentclass[pra,aps,twocolumn,superscriptaddress,showpacs]{revtex4}
\usepackage{graphicx} 
\usepackage{dcolumn} 
\usepackage{amssymb} 

\newcommand{\beqa}{\begin{eqnarray}}
\newcommand{\eeqa}{\end{eqnarray}}
\newcommand{\beq}{\begin{equation}}
\newcommand{\eeq}{\end{equation}}

\newcommand{\bfa}{\mathbf{a}}
\newcommand{\bfb}{\mathbf{b}}
\newcommand{\bfp}{\mathbf{p}}
\newcommand{\bfq}{\mathbf{q}}
\newcommand{\bfc}{\mathbf{c}}
\newcommand{\bfd}{\mathbf{d}}

\newcommand{\eipih}[2]{{\rm e}^{{\rm i}\frac{\pi}{2} #1\cdot #2}}
\begin{document} 
\title{Qubits in phase space: Wigner function approach to quantum 
error correction and the mean king problem}

\author{Juan Pablo \surname{Paz}} 
\affiliation{Departamento de F\'{\i}sica, FCEyN, UBA, Pabell\'on 1, 
Ciudad Universitaria, 1428 Buenos Aires, Argentina} 
\affiliation{Theoretical Division, LANL, MSB213, Los Alamos, NM 87545, USA} 
 
\author{Augusto Jos\'e \surname{Roncaglia}} 
\affiliation{Departamento de F\'{\i}sica, FCEyN, UBA, Pabell\'on 1, 
Ciudad Universitaria, 1428 Buenos Aires, 
Argentina} 
\affiliation{Theoretical Division, LANL, MSB213, Los Alamos, NM 87545, USA} 
 
\author{Marcos \surname{Saraceno}} 
\affiliation{Unidad de Actividad F\'{\i}sica, Tandar, CNEA, Buenos Aires, Argentina} 
\affiliation{Escuela de Ciencia y Tecnolog\'{\i}a, Univ. Nac. Gral. San Martin, Buenos
Aires, Argentina}

\date{\today}

\begin{abstract} 
We analyze and further develop a new method to represent the quantum state of 
a system of $n$ qubits in a phase space grid of $N\times N$ points (where $N=2^n$). 
The method, which  was recently proposed by Wootters and co--workers (Gibbons {\it et
al.}, quant-ph/0401155), is based on the use of the elements of the finite 
field $GF(2^n)$ to label the phase space axes. We present a self--contained 
overview of the method, we give new insights on some of its features and we 
apply it to 
investigate problems which are of interest for quantum information theory: We 
analyze the phase space representation of stabilizer states and quantum error 
correction codes and present a phase space solution to the so--called ``mean 
king problem''. 
\end{abstract} 
 
 
\maketitle 

\section{Introduction} 

Quantum mechanics can be formulated in phase space, the natural arena of classical 
physics. When doing this, quantum states are represented by phase space distributions 
with peculiar properties that distinguish them from their classical counterparts. 
The most common one is the Wigner function \cite{Wigner} which is a real--valued 
function that, unlike a genuine probability density, takes negative values
for generic quantum states. However, Wigner functions share an important property
with a classical probability density: when integrated along any phase space strip
(the region between any two parallel lines) is a positive number bounded between zero 
and one. The use of phase space methods has proved 
useful in the study of the quantum--classical transition and in the analysis of 
semi-classical properties of quantum systems. 

In the context of the recent surge of interest on quantum information the use of phase
space representation for quantum states and algorithms has been proposed \cite{MPSpra}. 
However, as opposed to what happens for continuous systems, the definition of Wigner functions
for the discrete case suffers from some ambiguities. 
In fact, there are various approaches available to generalize
the Wigner function for quantum systems with a finite--dimensional space of states.
Discrete versions of Wigner functions were first introduced in the context of studies
of semi-classical properties of classically chaotic systems by Hannay and Berry 
\cite{Hannay}. Some time later, Feynman \cite{Feynman} independently 
defined a phase space distribution for a spin-$1/2$ particle in a $2\times 2$ grid
(see also \cite{Cohen}). Wootters \cite{Wootters1} further developed this idea and defined 
the discrete version of the Wigner function for $N$--dimensional quantum systems when
$N$ is a prime number (see also \cite{Galetti}). In such case the phase space is 
an $N\times N$ grid. Wootters method can also be applied when $N$ is a composite 
number but in such case the Wigner function must be defined in a phase space grid which is 
the Cartesian product of the phase spaces corresponding to the prime factors of $N$. 
For the prime dimensional case, the phase space points $(q,p)$ have position and momentum
coordinates which are integers between $0$ and $N-1$. The fact that this set has the 
structure of a finite field is essential in implementing the method. A different approach
was followed by Cohendet \textit{et al.} \cite{Cohendet} who were able to define a Wigner 
representation for odd values of $N$ (not necessarily prime) using an $N\times N$ grid. 
Similarly, Leonhard \cite{Leonhardt} extended the idea for even values of $N$ but in this
case it was necessary to use a grid with $2N\times 2N$ points. Some of us proposed a similar
approach \cite{MPSpra} which can be used to define a discrete Wigner function for arbitrary
values of $N$ using a grid with $2N\times 2N$ points. In all these cases the phase space
coordinates $(q,p)$ are assumed to be integers between $0$ and $N-1$ (or $2N-1$). Arithmetic 
modulo $N$ is used when operating with such coordinates. The 
fact that for non--prime values of $N$ $Z_N$ is not a finite field is the cause of 
several peculiar and unpleasant properties of the phase space (for example, when defining
lines in phase space as solutions of linear equations one ends up with non--parallel lines
that intersect in more than one point). 
More recently Wootters \cite{Wootters2} proposed a different approach to define 
the discrete Wigner function that enables one to work on a 
phase space grid with $N\times N$ points when $N$ is the power of a prime number. 
The method, which was discussed in a comprehensive paper by Gibbons, Hoffman and Wootters
\cite{Wootters3}, heavily relies on the use of the elements of the finite field $GF(N)$
to label both phase space coordinates $(q,p)$. For various reasons this approach seems to 
be well suited to study some quantum information problems. Thus, quantum computers 
made out of $n$ qubits have a Hilbert space whose dimension $N=2^n$ is a power of a prime 
number. Moreover, operators made out of tensor products of Pauli operators acting on 
each qubit play an important role for quantum computers and have a central role in Wootters
phase space method (they represent phase space translation operators as discussed below). 

The use of the discrete Wigner functions in quantum information has been 
recently been proposed. Applications using Wigner functions in a $2N\times 2N$ 
grid include: the study of general properties of quantum states and the phase space 
representation of Grover's search algorithm \cite{MPSpra}, the phase space representation
of quantum teleportation of arbitrary dimensional systems \cite{Pazpra}, the study of 
quantum walk algorithms (where phase space methods seem to be particularly well suited)
\cite{Lopez}, the analysis of decoherence models with a natural phase space 
representation \cite{Bianucci}, applications to the use of quantum computers to study 
properties of classically chaotic maps \cite{Shepe} and the development of efficient  
techniques for quantum state tomography in phase space \cite{MPSnat,PRS03}. On the other hand, 
some attention was devoted to the use of the Wigner function originally defined by 
Wootters \cite{Wootters1} to analyze teleportation \cite{BuzekPra}. Following Wootters's 
more recent ideas \cite{Wootters2} a study of the nature of the set of quantum states
that have positive Wigner functions was presented by Galv\~{a}o \cite{Galvao} who also 
conjectured the existence of a connection between these states and the ones that can
be classically simulatable. 
However, it is still unclear if the use of phase space methods will help in expanding 
our understanding of some of the problems of quantum information. It is indeed possible
that the various definitions of Wigner function may end up proving to be useful to 
analyze different problems (indeed, much of the use of $2N\times 2N$ phase space representations
was done in the spirit of previous studies of semi-classical properties of quantum maps 
\cite{Hannay}). 

In this work we will present a new step towards exploring the use of
phase space techniques for quantum information using and expanding the recent work 
done by Wootters and co--workers \cite{Wootters1,Wootters2,Wootters3}.
As mentioned above, this method relies heavily on the fact that phase space coordinates
(position and momentum) are chosen to be elements of a finite Galois field. 
In this way it is simple to construct a phase space with geometric properties that are
similar to the usual phase space (i.e., where two lines that are not parallel have only 
one intersection, etc). Once the phase space arena is built in this way it is possible
to define ways to map sets of parallel lines (i.e., \textit{striations}) to bases of the 
Hilbert space, and to show that the bases associated with different striations turn out to
be mutually unbiased. In our paper we will discuss some interesting properties of this 
phase space for systems made out of qubits. We will analyse a few problems where these 
phase space tools could be naturally applicable. 

The paper is organized as follows. In Section II we review, in a self--contained way, 
the main properties of the discrete phase space. Here, we also show how to reduce the 
arbitrariness in the association between lines and states (fixing the \textit{quantum 
net}) by imposing additional symmetries to the phase space structure. In Section III 
we show how to define the discrete Wigner function and discuss some of its most important
properties. In Section IV we analyze quantum information problems using this discrete 
Wigner function: First we analyze properties of stabilizer states \cite{Gottesman},
which are eigenstates of translation operators. 
Finally, we show a phase space solution to the \textit{mean king problem} \cite{Vaidman, Aharonov}. 
In Section V we analyze some general properties of the phase space representation of stabilizer 
states, we discuss possible future directions and summarize our results.

\section{Discrete phase space for a system of $n$ qubits} 
\subsection{Phase space coordinates}

In this section we will present a self contained review of a recent method to 
define a phase space for a system of $n$ qubits. This discrete phase space is the 
natural arena in which a discrete Wigner function can be introduced. 
Our treatment is based on the ideas discussed first by Wootters \cite{Wootters1} 
and developed by Gibbons, Hoffman and Wootters 
\cite{Wootters2,Wootters3}. We will concentrate on systems of qubits but 
the method can be extended to systems with Hilbert space with a dimension that 
is a power of a prime number. 

For a system with $n$ qubits we introduce a phase space grid with $N\times N$ points
($N=2^n$ denotes the dimension of the Hilbert space). 
We label each phase space point with its ``position'' and ``momentum'' coordinates $(q,p)$, 
where both $q$ and $p$ are elements of the field $GF(2^n)$. 
The elements of this field can be thought of as being the $N=2^n$ $n$--tuples with 
binary (i.e., $0$
or $1$) entries. The sum in $GF(2^n)$ is defined as the bitwise (mod $2$) addition of the 
respective $n$--tuples. Elements of $GF(2^n)$ can also be thought of as polynomials of degree
$n-1$ with binary coefficients. The product in $GF(2^n)$ is defined as the product of the 
corresponding polynomials modulo a {\it primitive} polynomial (this is a polynomial of degree
$n$ which is irreducible, i.e. it cannot be factored as the product of two polynomials of 
lower degree). The product in $GF(2^n)$ is such that the set formed by all the elements
of $GF(2^n)$ excluding the zero is a cyclic group of order $2^n-1$ with the multiplication 
in $GF(2^n)$ as the group operation. Thus, $GF(2^n)$ contains the zero element 
and the $2^n-1$ different powers of a primitive element $\omega$:
$\{0,1,\omega,\omega^2,\ldots,\omega^{2^n-2}\}$ (note that $\omega^{2^n-1}=1$).
This establishes a natural order for the field elements, and we use this order to
label the axes.  Using the 
primitive polynomial it is clear that any power $\omega^k$ with $k\ge n$ can be expressed
as a linear combination of powers of degree lower than $n$. This defines the addition 
rule in 
the field. For example, for $n=3$ (three qubits) the polynomial $\pi(x)=x^3+x^2+1$ can be 
chosen as primitive. Then, the field is formed by the following elements 
$GF(2^3)=\{0,1,\omega, 
\omega^2, 1+\omega^2, 1+\omega+\omega^2, 1+\omega, \omega+\omega^2\}$ (the last four elements
in the list are the powers $\omega^k$ with $k=3,4,5,6$). An ordered set of $n$ elements
of the field, $E=\{e_0,\ldots,e_{n-1}\}$, is a basis if every element of 
$GF(2^n)$ can be written as the linear combination
\begin{equation}
x=\sum_{i=0}^{n-1} x_ie_i.\nonumber
\end{equation}
In this equation all the coefficients $x_i$ are elements of $GF(2)$ (i.e., they are 
either $0$ or $1$). The dual basis of $E$ can be introduced as follows: The 
trace of any field element $x$ is defined as
\begin{equation}
\text{tr}(x)=x+x^2+x^{2^2}+\ldots+x^{2^{n-1}}.\label{trace}
\end{equation}
This operation is linear, which follows from the fact that for every pair of 
elements of $GF(2^n)$ the square of their sum is equal to the sum of their squares:
i.e., $(x_1+x_2)^2=x_1^2+x_2^2$. Moreover, it can be shown that the trace (\ref{trace})
takes values in $GF(2)$. Using the above trace one can define a dual for every 
field basis: Thus, it turns out that for every basis $E$ there 
is a unique basis $\bar E$ satisfying that ${\rm tr}(\bar e_ie_j)=\delta_{ij}$. 
The basis $\bar E$ is the dual of $E$.

There is a very useful representation for $GF(2^n)$ in terms of $n\times n$ matrices
with binary coefficients. It is worth pointing out here the main properties of this
representation, which are discussed in more detail in the Appendix. Given a primitive
polynomial $\pi(x)$ for the field $GF(2^n)$ one can choose its companion matrix $M$ 
as the primitive element for the matrix representation (such matrix is defined in the
Appendix and is such that $\pi(M)=0$ and $M^{2^n-1}=1$). The first $n$ powers of the 
matrix $M$ can be chosen as the canonical basis for the field and any field element
is then written as $x=\sum_{i=0}^{n-1}x_iM^i$ with the $n$ binary coefficients 
$x_i$ being the coordinates representing $x$ as an $n$--tuple. The virtue of the
matrix representation of $GF(2^n)$ is that both the sum and the product in the 
field correspond directly to ordinary operations between matrices. Also, the 
trace of any element, defined in (\ref{trace}) is nothing but the trace of the 
corresponding matrix. 
These are the basic mathematical elements about $GF(2^n)$ that are necessary 
to build the phase space arena to represent the quantum states of $n$ qubits. 

\subsection{Lines and translations in phase space}

In the $N\times N$ phase space grid we define a line as the set of phase space points
satisfying a linear equation $aq+bp=c$ (where all elements and operations 
appearing in this equation belong to $GF(2^n)$). Two lines are parallel if they do not 
intersect, or equivalently if they satisfy the equations $aq+bp=c$ and $aq+bp=c'$ with 
$c\neq c'$. 
Lines in this phase space grid have rather natural properties: 
Given a line it is always possible to find $N-1$ other lines that are parallel to it. A set of 
$N$ parallel lines form a {\it striation} of the phase space (i.e., a striation is 
a {\it foliation} of the grid with parallel lines). It is easy 
to show that there are $N+1$ different striations of the phase space. Also, because $N$ 
is a power of a prime number, two lines that are not parallel intersect only at one point.

\begin{figure} 
\includegraphics[width=5cm]{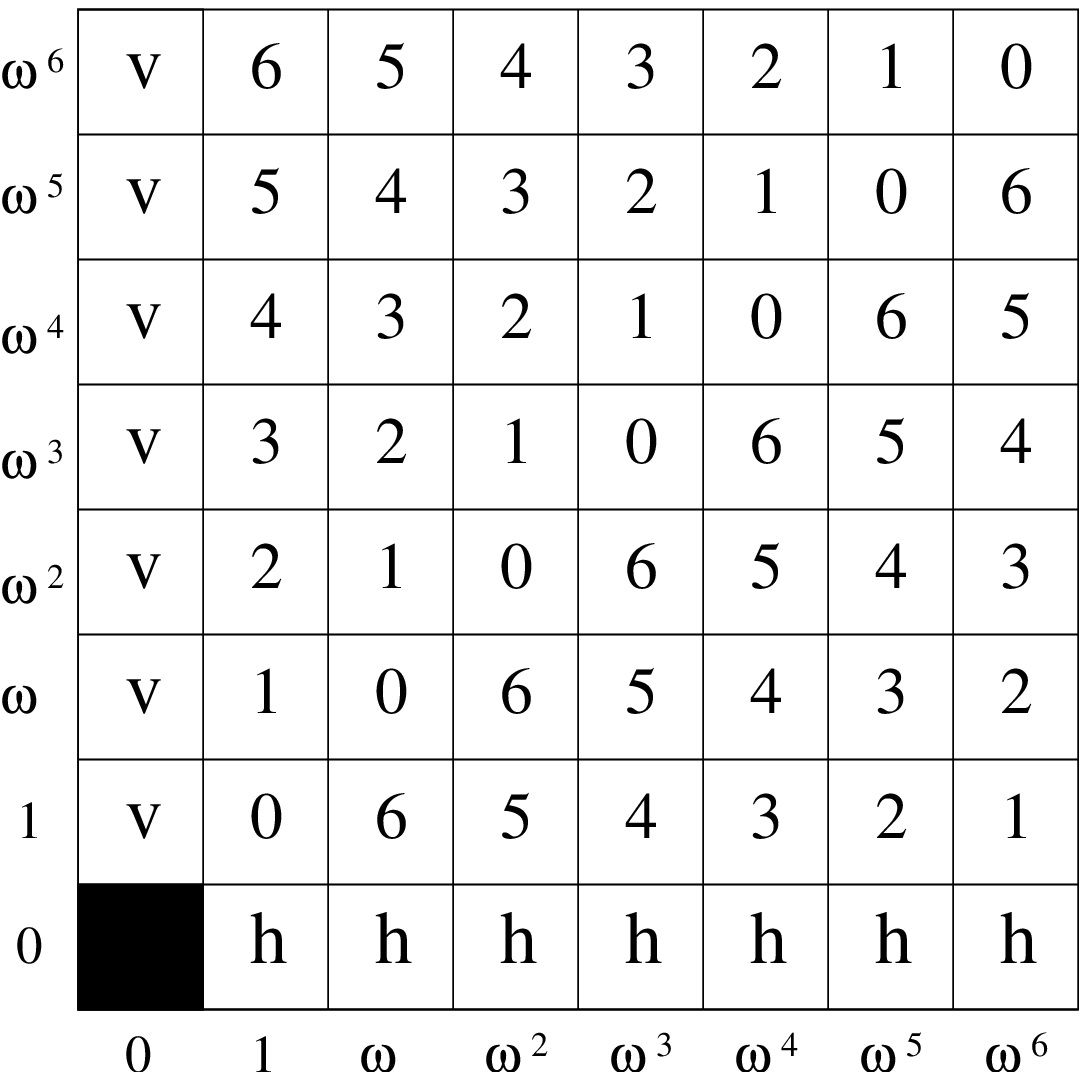} 
\caption{Lines passing through the origin (rays) of phase space for $3$ qubits. 
All rays apart from the vertical and horizontal ones (labeled with $h$ and $v$ in
the plot) satisfy the equation $p=\omega^j q$ (the power $j$ defines the slope of 
each curve and its value is displayed inside each cell). The total number of lines 
in this case is $N+1=9$. The position and momentum axes are labeled by elements of the 
field $GF(2^3)$.} 
\label{grid} 
\end{figure}

A phase space translation by an amount $\alpha_0=(q_0,p_0)$ will transform any 
phase space point $\alpha=(q,p)$ into $\tau_{\alpha_0}\alpha=\alpha+\alpha_0$. Lines in phase
space are invariant under some translations. Consider the line containing the origin
and the point $(q,p)$. This line is formed by the points  
$(sq,sp)$, where $s$ is a parameter that ranges over all the elements of $GF(2^n)$. The 
striation containing this line remains invariant under any translation of the 
form $\tau_{(sq,sp)}$. We will take as a reference line for each striation the line passing
through the origin, specified by the equation $aq+bp=0$, which we call a 
ray \cite{Wootters3}. Notice that if 
the axes are labeled with the powers of a primitive element these rays appear on the main 
diagonals and look ``parallel" in the ordinary sense, wrapping periodically on a torus
of $2^n-1$ periodicity. The $N+1$ rays for three qubits are displayed in 
Fig.\ref{grid}.

\subsection{Translation operators }

The association of a translation operator $T_\alpha$ with a phase space point $\alpha=(q,p)$ 
should be done in such a way as to preserve the additive structure of the field. Moreover,
following \cite{Wootters1,Wootters2,Wootters3}, we require that for a system of $n$
qubits translations should act independently on each qubit, thus preserving the tensor product
structure of the Hilbert space (this is a nontrivial assumption which is 
not satisfied in other constructions \cite{MPSpra}).  At the 
level of individual qubits position and momentum translations are identified with 
Pauli operators $X_i$ and $Z_i$. Operators that satisfy these requirements are
\beq
T(\bfq,\bfp)= \prod_{i=0}^{n-1} {X_i}^{q_i}{Z_i}^{p_i}\eipih{q_i}{p_i}
               \equiv X^\bfq Z^\bfp \eipih{\bfq}{\bfp},
\label{translations}
\eeq
where we denote $ (\bfq,\bfp) $ the binary strings $(q_0 \ldots q_{n-1} , p_0\ldots p_{n-1})$. 
The phase $\bfq\cdot\bfp= \sum q_i p_i $ is added to
make the operators both unitary and hermitian. It is simple to show that the 
factor $\exp(i\pi\bfq\cdot\bfp/2)$ can take values $\pm 1$, $\pm i$ and that
$T(\bfq,{\bf 0})=X^\bfq,~T({\bf 0},\bfp)=Z^\bfp,~T(\bfq,\bfq)=Y^\bfq$.
The set of $N^2$ operators $T(\bfa,\bfb)$ is an orthogonal basis of the space of operators
since $ {\rm Tr} (T(\bfa,\bfb )T(\bfc, \bfd ))=N\delta_{(\bfa,\bfc)} \delta_{(\bfb,\bfd)}$. 

Associating a translation $T(\bfq,\bfp)$ with a phase space point $\alpha=(q,p)$
requires a mapping between the field elements $q$ and $p$ and the binary strings $\bfq$ and $\bfp$.
The most natural such mapping is to use the binary string $\bfq$ ($\bfp$) 
formed by the coefficients of the expansion of the field elements $q$ ($p$) in a 
given field basis. The field basis is arbitrary and could in principle be different for 
position and momentum. However, we will show below that the consistency of the phase 
space structure implies that once we choose a basis to expand the position $q$, there are strong 
constraints on the basis we could use to expand the momentum $p$. Before doing this, it 
is worth noticing two simple properties of the translation operators defined above:  
The composition law for the translations is 
\beq
T(\bfa,\bfb)T(\bfq,\bfp)= \pm T(\bfa + \bfq, \bfb+\bfp)
      \exp ({\rm i}\frac{\pi}{2}(\bfb\cdot\bfq-\bfa\cdot\bfp)),
\label{composition}
\eeq
which leads to the commutator 
\beq
[T(\bfa,\bfb),T(\bfq,\bfp)] =\pm 
    2i \sin\frac{\pi}{2}(\bfb\cdot\bfq-\bfa\cdot\bfp)T(\bfa + \bfq, \bfb+\bfp).
 \label{commutator}
\eeq
Thus two translations commute {\it iff} $ \bfa \cdot \bfp-\bfb\cdot\bfq = 0 \pmod{2}$.
 
Using the results in the Appendix we can then construct commuting sets of translations
as follows: consider the set of translations
\begin{equation}
T(\bfa M^j ,\bfb \tilde{M}^j) \qquad\qquad j=0,2^n-2,
\label{commutingsets}
\end{equation}
where $M$ is the companion matrix of the primitive polynomial used to construct the
field and $\tilde M$ is its transpose. Using Eq. (\ref{commutator}) it is
easy to show that these operators commute. These operators, together with $T({\bf 0},{\bf 0})$ 
form a complete set of $N$ commuting translations whose common set of eigenvectors constitute
a basis for the Hilbert space. Using this procedure we can partition the set of all 
$N^2$ translation operators into $N+1$ sub--sets of 
commuting translations. Each of these sub--sets contains the identity and other $N-1$ 
operators which are of the form (\ref{commutingsets}). Thus, each of these $N+1$ 
sub--sets can be associated with different strings $(\bfa,\bfb)$. The following is a 
simple way to choose the $N+1$ strings to define the corresponding sets: We can first 
define two sets associated with the strings $(\bf1, \bf0)$ and $(\bf0, \bf1)$. 
Then, the remaining $N-1$ sets can be associated with strings of the form 
$(\bf1, \bfb)$ for all $\bfb\neq\bf0$. Here, and below, we use the notation
${\bf1}=(10\ldots 0)$ and ${\bf0}=(00\ldots 0)$. We remark that this is a very 
simple way to find the $N+1$ commuting sets of translations defining the corresponding
MUBs. The above choice of binary strings $\bfa$ and $\bfb$ is arbitrary but it is easy
to see that different choices of such strings would lead to the same $N+1$ sets. 
There are two extra freedoms that one may exploit. First, one can change the 
companion matrix $M$ by choosing a different primitive polynomial. 
Second, one can define the translation operators in a different way than the one given
in eq. (\ref{translations}). For example, for each qubit we could interchange the 
three Pauli operators. In this way one obtains a different partition of the Pauli group 
into $N+1$ commuting sets.

\subsection{Association between lines and states}

With the above tools we are ready to address the first fundamental point made 
by Wootters in defining the phase space. The key to this construction is to establish
a one to one correspondence between every phase space line $\lambda$ and a state in 
Hilbert space. 
We will describe how to associate a pure state with the line $\lambda$ 
(or, analogously, define a mapping between every line $\lambda$ and a rank one projection 
operator $P(\lambda)$).  
Following Wootters this mapping is defined by imposing a natural geometric constraint: we  
require that the mapping $P(\lambda)$ should be such that the state associated with the 
translated line $\tau_\alpha\lambda$ should be obtained from $P(\lambda)$ by applying a 
translation operator $T_\alpha$, i.e.
\begin{equation}
P(\tau_\alpha\lambda)=T_\alpha P(\lambda)T^\dagger_\alpha.\label{translatelines}
\end{equation}

This condition enforces the validity of two important results. The first result 
following from (\ref{translatelines}) is that lines belonging to the same 
striation are associated with orthogonal states (i.e., a striation is 
associated with an orthonormal basis of the Hilbert space). The reason why this
is the case is the following: Consider the striation containing the ray formed by the
points $(sa,sb)$ where $a$ and $b$ are fixed and $s$ ranges over all elements of $GF(2^n)$. 
All lines in this striation are invariant under translations $\tau_\alpha$ with 
$\alpha=(ta,tb), \forall t\in GF(2^n)$. Therefore, as $\tau_\alpha\lambda=\lambda$, 
equation (\ref{translatelines}) implies that 
all the operators $T_\alpha$ must commute with the projectors $P(\lambda)$. Thus, 
this implies that the translation operators associated with every point in 
the ray $(sa,sb)$ must form a commuting set and, moreover, that the states associated
with the striation containing the ray must be the common eigenstates of the translations. 

The second result following from (\ref{translatelines}) is that the bases associated 
with the $N+1$ different striations are mutually unbiased. In fact, one of the virtues 
of the phase space structure based on finite fields is that it establishes a clear 
connection between phase space striations and mutually unbiased bases (MUB). This 
concept is interesting on its own and has been studied in detail  
\cite{Ivanovic,WFields,Bandyo,LBZ}. Two 
bases sets $\{|\phi_j\rangle, \ j=1,\ldots,N\}$ and $\{|\psi_k\rangle,\ k=1,\ldots,N\}$ 
are MUB if and only if $|\langle\phi_j|\psi_k\rangle|^2=1/N$ for all values of $j$ and $k$. 
It has been shown that $N+1$ MUB exist when $N$ is the power of a prime 
number \cite{WFields}. The interest in MUB is strongly connected with the problem 
of state tomography. Thus, one can show that the most efficient way of 
completely determining a quantum state is by making von Neumann measurements on the 
$N+1$ MUB \cite{WFields}. Notably, Wootters phase space construction is itself a 
relatively simple method for explicitly constructing the $N+1$ sets of MUB. 

The association between striations and MUB also appears in a transparent 
manner: As we mentioned above, each striation should be associated with a set 
of $N-1$ nontrivial commuting translations. As we have a total of $N^2-1$ nontrivial
translation operators (we do not include the identity) we can split them into 
$N+1$ disjoint sets containing $N-1$ operators each. In fact, this was done explicitly
above. All operators within each set commute and the operators belonging to different 
sets are orthogonal. When these conditions are met a powerful theorem proved by 
Bandhyopadyay \textit{et al.} holds 
\cite{Bandyo}: If a set of $N^2-1$ traceless operators can be split into $N+1$ orthogonal 
sets of commuting operators then the eigenbases associated with these sets are mutually 
unbiased. 

It is very important to notice that condition (\ref{translatelines}) also implies that 
the mapping between field elements $q, p$ and binary strings $\bfq, \bfp$
cannot be arbitrary. Let us consider the ray formed by the points $(sa,sb)$ (which 
satisfy the equation $bq+ap=0$). As described above, the points in this ray are 
associated with the translation operators $T(\bfa M^j,\bfb \tilde M^j)$. Consider in 
particular the horizontal ray $p=0$, which contains all points $(sa,0)$. Apart from
the origin $(0,0)$ all the points in this ray can be ordered according to the powers
of the generating element $\omega$ since they can be expressed as $(\omega^j ,0)$. 
Therefore, the binary strings corresponding to these points should be of the form 
$(\bfa M^j,\bf0)$, and the origin. Choosing $\bfa={\bf1}=(1,0,\ldots,0)$ the 
matrix $M$ determine the remaining
binary strings associated with the position axis of phase space. These binary strings are 
nothing but the coefficient of the expansion of the field element $q$ in the canonical 
basis formed by the first $n$ powers of $\omega$ (of course, choosing a different binary 
string $\bfa$ for the first element is tantamount to a change of basis). 
The binary strings associated with the momentum axis can also be determined in this 
way. Consider the vertical ray $q=0$ which contains the origin and all the points of 
the form $(0,\omega^j)$. These points are associated with the binary strings 
$({\bf0},\bfb \tilde M^j)$, and the origin. Again, the choice of the binary string associated with the first
point in the ray is arbitrary and we can choose it to be $\bfb=\bf1$. The subsequent 
binary strings associated with the momentum axis are therefore determined as $\bfp={\bf1}
\tilde M^j$. 

It is interesting to notice that the matrices $M$ and $\tilde M$ have dual roles: 
While the powers of $M$ can be used to find the 
coordinates of field elements in the canonical basis, the powers of $\tilde M$ 
enable us to 
find the coordinates of field elements in the dual basis. Therefore, our previous argument
implies that the binary strings associated with the momentum axis of phase space are the 
components of the field elements in a basis which is a multiple of the dual of the 
canonical
basis (the freedom in choosing $\bfb$ implies that the basis that should be used for the 
momentum axis is a multiple of the dual and not simply the dual). 
 
As we described above, the condition (\ref{translatelines}) is crucial in establishing
a relation between a basis set in Hilbert space and a striation of phase space. 
However, it does not tell us what specific state one should associate
with each line of the striation. In fact, this association is entirely arbitrary. To 
completely define the phase space structure one should provide such relation between 
states and lines. In this way, using the terminology employed by Wootters \textit{et al.}
\cite{Wootters3} one defines a ``quantum net''. Thus, 
there are many quantum nets allowed by equation (\ref{translatelines}). 

One can calculate the number of allowed quantum nets as follows. Let us assume, for 
simplicity,
that the association between states and vertical or horizontal lines is fixed. There are
$N-1$ remaining striations and each of the corresponding rays can be associated with any 
of the $N$ states of the corresponding MUB. Therefore, there are $N^{N-1}$ quantum nets
consistent with the above constraints 
and each one will lead to a different Wigner function as described below.

\subsection{Fixing the {\sl quantum net}}

To specify a quantum net we need to make explicit the connection between every line
in a striation and a state in a basis of the Hilbert space. For this purpose we only need
to assign a state to the ray of the striation. Thus, once we do that we can obtain
the states associated with all the other lines of the striation 
by applying the translation operator that maps the ray to the line.  
We will adopt a labeling scheme 
for the rays as in Fig. \ref{grid}. The ray $\lambda$ is the line $p=\omega^\lambda q$ and
we label separately the horizontal $(p=0)$ and the vertical $(q=0)$ rays as $\lambda=h,v$.

It is useful to notice that a rank one projector onto an eigenstate of the translation 
operators associated with a ray can be constructed as follows: The first $n$ points in each 
ray (besides the origin) correspond to a set of $n$ translation operators which can be 
considered as the generators of the set of translations associated with the ray. 
In fact any set of $n$ non--trivial translations belonging to the ray could be used as generators 
since the rest of the operators are obtained from all possible products of the $n$ 
generators. Then the generators for ray $\lambda$ are the operators 
\beq
G^{(\lambda=j)}_k = T({\bf 1}M^k,{\bf 1}\tilde{M}^{(k+j)}),\qquad k=0,\ldots,n-1,
\nonumber
\eeq  
where, again, ${\bf 1}$ is the binary string $(10\ldots 0)$. 

A possible, and simple, way to fix the quantum net is by associating each ray with 
the state which is an 
eigenstate of all the generators with eigenvalue $+1$. Recall that the translation 
operators are hermitian and unitary and therefore have eigenvalues $\pm 1$. Therefore 
this projector is 
\beq
 P^{(\lambda)}_0 =\frac{1}{2^n}\prod_{k=0}^{n-1} ({\mathbb I}+ G^{(\lambda)}_k).
 \label{proyector}
\eeq
The projectors on the other lines of the striation are obtained by translating 
the above one. In this way we have arbitrarily fixed the quantum net. Notice 
that the association of a given state to each ray is an independent process. 
Instead, once the state associated with a ray is chosen, the states to be 
assigned to the rest of the striation are completely determined  
by the covariance requirement under translations imposed in Eq. (\ref{translatelines}).
 
However, there operations which are not translations but map lines
into other lines, thus leaving the phase space invariant: these are the unit determinant
linear transformations. In the usual continuous case these are the symplectic 
transformations, and the usual Wigner function is also covariant under these 
more general phase space maps. It is then natural to try to reduce the number of 
possible quantum nets by imposing additional 
symmetries. Unfortunately, as discussed in \cite{Wootters3}, it is not possible to find a 
faithful unitary representation for arbitrary linear, unit determinant transformation. 
However, there is an important transformation for which this is indeed possible: 
the rescaling (or squeezing) $u_\omega$ defined as
\beq
u_\omega(q,p)=(\omega q, \omega^{-1}p).
\label{rescaling}
\eeq
In the Appendix we describe a general quantum circuit, made out of simple swaps and 
controlled nots, defining the unitary operator $U_\omega$ that represents the 
transformation $u_\omega$ in Hilbert space. This operator is such that
\beq
U_\omega X^\bfa U^\dagger_\omega=X^{\bfa M}\qquad
U_\omega Z^\bfb U^\dagger_\omega=Z^{\bfb \tilde M^{-1}}. 
\nonumber
\eeq
The operator $u_\omega$ maps rays into rays: while leaving the vertical and horizontal rays
unchanged it cycles through all the diagonal ones. In fact, using the notation introduced
above, for every ray that is not vertical nor horizontal the squeezing 
transformation is such that $u_\omega\lambda=\lambda-2 \pmod{N-1}$. 
The existence of this transformation can be used to reduce the arbitrariness in defining
the quantum net. Thus, once we fix the state to be associated to a diagonal ray, 
say $P^{(\lambda=0)}_0$, then the states associated with all the 
other rays are not independent any  more. In fact, they are fixed by the covariance 
requirement
\beq
P(u_\omega\lambda)=U_\omega P(\lambda) U^\dagger_\omega.
\eeq
If this condition is imposed, the arbitrariness in the quantum net is greatly reduced 
from $N^{N-1}$ to $N$.

The operator $U_\omega$ has another very useful property. As discussed above, $U_\omega$ 
provides the change of basis from the one associated with the striation $\lambda$ 
onto the one corresponding to the striation $\lambda-2 \pmod{N-1}$. Suppose that we
are interested in explicitly obtaining the states of all $N-1$ MUB associated with  
the striations which are not vertical nor horizontal. For this we would only need
to find the states associated with one striation, say the one corresponding to 
$\lambda=0$ whose generators are of the 
form $G^{(0)}_k = T({\bf 1}M^k,{\bf 1}\tilde{M}^{k})$. Once we find the 
common eigenstates of these generators we can easily construct the states of 
all the remaining $N-2$ MUB by simply applying the operator $U_\omega$. The 
simplicity of the circuit representing
$U_\omega$ makes this a very efficient method to obtain the states of all MUB.  

\subsection{Example 1: Phase space for $n=2$ qubits} 
 
\begin{figure}  
\includegraphics[width=5cm]{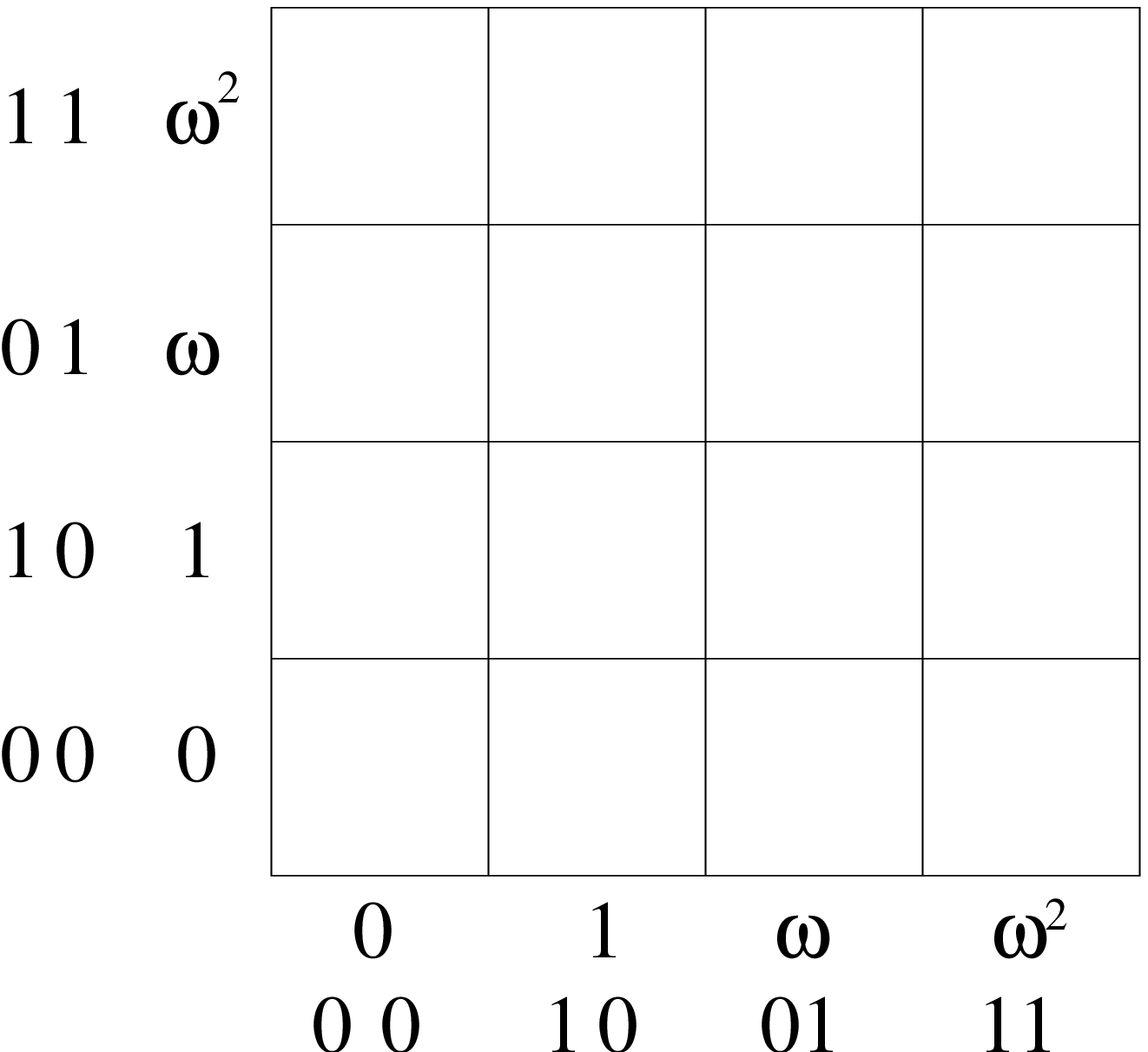}  
\caption{Phase space for a system of two qubits. Two labellings appear in the axes:
One labeling associates position and momentum with elements of the field $GF(4)$. 
The other labeling associates them with binary strings (which can be directly mapped
onto quantum states).}
\label{n=2}  
\end{figure}

\begin{figure}  
\includegraphics[width=5cm]{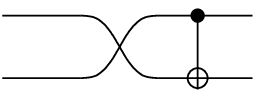}  
\caption{Quantum circuit representing the operator $U_\omega$ for two qubits. This
operator acts on the states associated with vertical lines by ``moving them to the right''
(i.e. mapping the state associated with the line $q$ onto the state associated
with the line $\omega q$). Similarly, this operator acts on states associated with 
horizontal lines by ``moving them downwards''. The states associated with 
the lines $q=0$ and $p=0$ remain invariant.}
\label{circuit-n=2}  
\end{figure}

We will show, as an example, how to construct the discrete phase space for systems of $2$  
qubits ($N=4$) using the procedure described above.  This case was already discussed in 
detail by Wootters in \cite{Wootters2,Wootters3} but we include it here for completeness. 
The field $GF(2^2)$ has four elements. The primitive 
polynomial can be chosen as $\pi(x)=x^2+x+1$. The primitive element $\omega$ is 
the root of such polynomial (which clearly does not belong to $GF(2)$). The field
has then four elements: the zero and the three powers of $\omega$, $GF(2^2)=\{0,1,\omega,
1+\omega\}$. The field has a canonical basis $e_i=\{1,\omega\}$ 
whose dual is $\bar e_i=\{1+\omega,1\}$. As we mentioned above, the basis $f_i$ to 
expand the momentum can be chosen as a multiple of $\bar e_i$. It
is convenient to chose $f_i=\omega \bar e_i$ since in this case both bases turn
out to be the same, i.e. $f_i=e_i$. The phase space coordinates are the ones seen
in Fig. \ref{n=2}. The generators of the horizontal, vertical and $\lambda=0$ 
striation are:
\begin{eqnarray}
G^{(v)}_0 &=& T(00,10),\quad G^{(v)}_1=T(00,01),\nonumber\\
G^{(h)}_0 &=& T(10,00),\quad G^{(h)}_1=T(01,00),\nonumber\\
G^{(0)}_0 &=& T(10,10),\quad G^{(0)}_1=T(01,01).\nonumber
\end{eqnarray}
To the vertical, horizontal and $\lambda=0$ ray we associate the eigenstate
with eigenvalue $+1$ of the corresponding generators. The states associated 
with the two other rays $\lambda=1,2$ can be obtained from the above by 
applying the operator $U_\omega$. For the system of two qubits this operator
is represented by the quantum circuit shown in Fig. \ref{circuit-n=2}. 
As mentioned above, if we apply this operator to the eigenstates of $G^{(0)}_0=Y_0$
and $G^{(0)}_1=Y_1$ we obtain the elements of the MUB corresponding to the 
striation $\lambda=2$. Acting once more with this operator we obtain the 
basis corresponding to $\lambda=1$. These bases are known as ``Belle'' and ``Beau''
\cite{Wootters2} and this is a rather simple way to determine their states. 


\subsection{Example 2: Phase space for $n=3$ qubits} 
 
\begin{figure}  
\includegraphics[width=5cm]{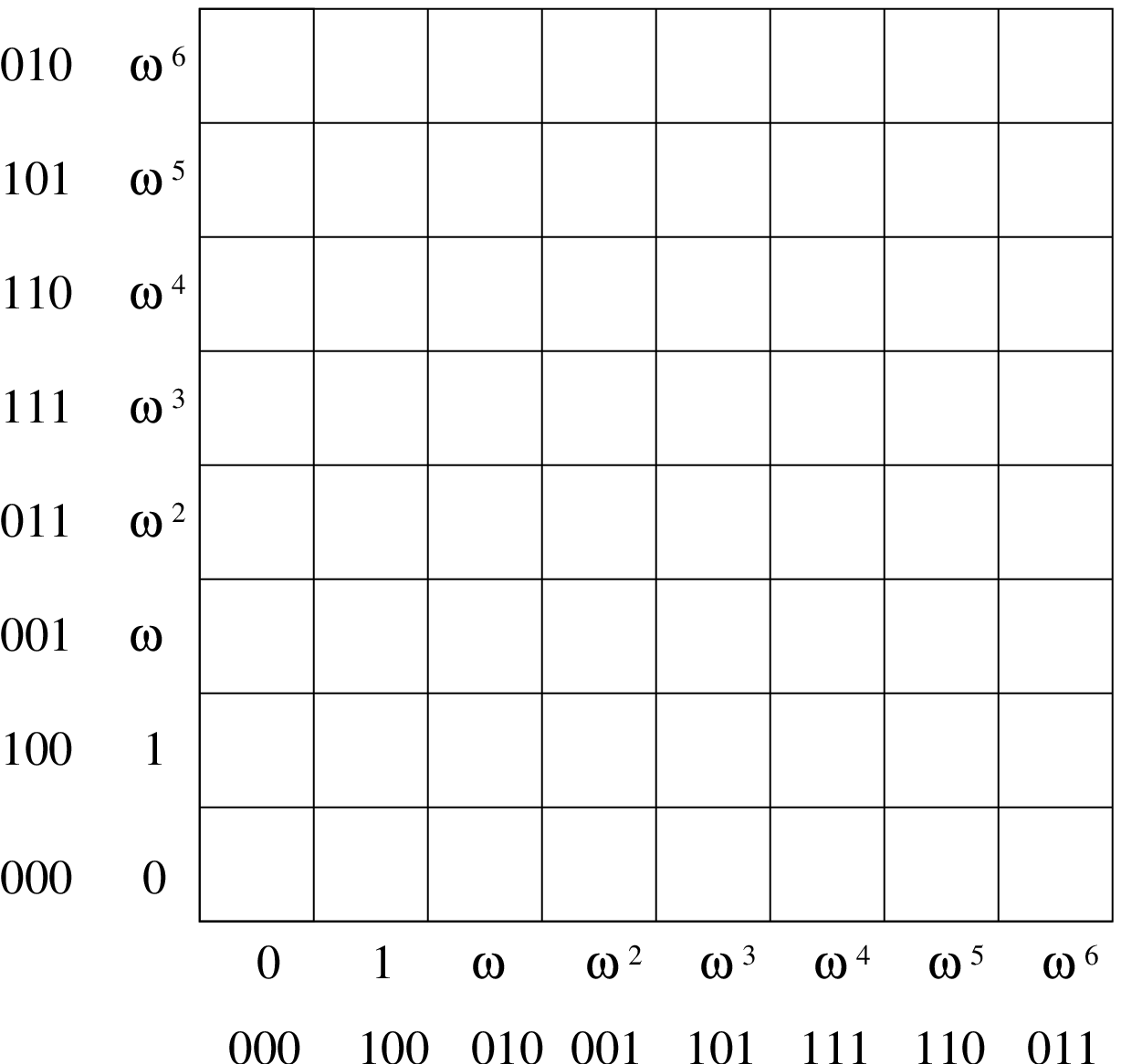}  
\caption{Phase space for 3-qubits systems. }  
\label{n=3}  
\end{figure}

\begin{figure}  
\includegraphics[width=5cm]{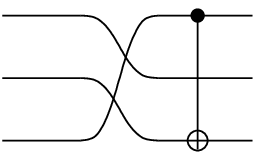}  
\caption{Quantum circuit representing the operator $U_\omega$ for three qubits. }
\label{circuit-n=3}  
\end{figure}

For $n=3$ position and momentum labels take values in the field $GF(2^3)$. 
As mentioned above, the primitive polynomial can be taken to be $\pi(x)=x^3+x^2+1$
and the field consists of the following elements $GF(2^3)=\{0,1,\omega,\omega^2,
1+\omega^2,1+\omega+\omega^2,1+\omega,\omega+\omega^2\}$. The canonical basis
$e_i=\{1,\omega,\omega^2\}$ has a dual given by $\bar e_i=(\omega^4,\omega^3,\omega^5)$. 
If we consider the basis $f_i=\omega^3 e_i$ we find that $f_i=\{1,\omega^6,\omega\}$. 
It is interesting to notice that in this case it is not possible to use $f_i=e_i$ as
we did in the simpler $n=2$ case. The phase space coordinates are shown in 
Fig. \ref{n=3}. The generators for the horizontal, vertical and $\lambda=0$ rays
are (we are using the ordering given in the Appendix in Table \ref{table1})
\begin{eqnarray}
G^{(h)}_0&=&T(100,000),\quad G^{(h)}_1=T(010,000),\nonumber\\
G^{(h)}_2&=&T(001,000),\nonumber\\
G^{(v)}_0&=&T(000,100),\quad G^{(v)}_1=T(000,001),\nonumber\\
G^{(v)}_2&=&T(000,011),\nonumber\\
G^{(0)}_0&=&T(100,100),\quad G^{(0)}_1=T(010,001),\nonumber\\
G^{(0)}_2&=&T(001,011).\nonumber
\end{eqnarray}
Again, the states associated with these rays can be explicitly constructed
using expressions like (\ref{proyector}). The states corresponding to the 
other rays are obtained by applying the operator $U_\omega$, that is now 
represented by the quantum circuit shown in Fig. \ref{circuit-n=3}.

\section{Discrete Wigner function for a system of $n$ qubits}

We will define here the discrete Wigner function using the phase space structure
that we introduced in the previous section. The Wigner function provides a 
representation of any quantum state in phase space, The definition we use will be 
such that the discrete Wigner function has the same three crucial properties than
its continuous counterpart: P1) The Wigner function $W(q,p)$ is real valued. 
P2) The Wigner function provides a complete description of the state and is such that the
inner product between any pair of states $\rho_1$ and $\rho_2$ can be obtained as
$\text{Tr}(\rho_1\rho_2)= N\sum_{q,p} W_1(q,p)W_2(q,p)$. P3) The sum of values of the Wigner
function along any line in phase space is equal to the probability of detecting the 
state associated with the line. 

It is interesting to notice that once we define a phase space structure (or a 
``quantum net'') the Wigner function is uniquely determined by the condition P3). 
This can be seen as follows: Consider the point $\alpha=(q,p)$ and all the lines
$\lambda$ that cross this point. These lines only intersect once, precisely at
the point $\alpha$. Therefore, the sum of the values of the Wigner function 
over all these lines is
\begin{equation}
\sum_{\beta\in\lambda/\ \alpha\in\lambda}W(\beta)=N W(\alpha)+\sum_\beta W(\beta).\label{sumW}
\end{equation}
The condition P3) implies obviously that the sum of values of the Wigner function over all
phase space should be unity. On the other hand, this condition also implies that
the left hand side of (\ref{sumW}) should be equal to the sum of expectation values 
of the projection operators associated with the lines $\lambda$. Therefore, we obtain 
that P3) implies that the Wigner function at any phase space point is
\begin{equation}
W(\alpha)={1\over N}(\sum_{\lambda/\ \alpha\in\lambda}\text{Tr}(\rho P(\lambda))-1).\label{walpha}
\end{equation}
This equation can be considered to be the definition of the discrete Wigner function. 
Equivalently, the Wigner function can be seen to be the expectation value 
of a ``phase space point operator'' $A(\alpha)$:
\begin{equation}
W(\alpha)=\text{Tr}(\rho A(\alpha)).\label{Wdef}
\end{equation}
According to (\ref{walpha}) the phase space point operator should be defined as:
\begin{equation}
A(\alpha)={1\over N}(\sum_{\lambda/\ \alpha\in\lambda}P(\lambda)-\mathbb{I}).\label{aalpha}
\end{equation}
The phase space point operators have several noticeable properties: From their 
definition it is clear that they are hermitian. Also, we can show that these 
operators form a complete basis of the space of operators. In fact, the Schmidt
inner product between any such operators is 
\begin{equation}
\text{Tr}(A(q,p))A(q',p'))={1\over N}\delta_{q,q'}\delta_{p,p'}, \label{innerprod}
\end{equation}
where $\delta_{x,x'}$ denotes the Kronecker--delta symbol (the validity of this 
equation follows directly from (\ref{aalpha}) using the properties of the 
phase space lines discussed above \cite{Wootters3}). 

Expressing the Wigner function in terms of the phase space point operators 
is useful since it makes clear the reason why properties P1)--P3) are satisfied: 
P1) is a consequence of the fact that $A(q,p)$ are hermitian operators. 
P2) follows from the fact that $A(q,p)$ form a complete orthonormal basis of 
the space of operators. P3) follows from the fact that the sum of $A(q,p)$ over
a line is nothing but the projection operator onto the state associated with 
that line. 

Using the fundamental equation (\ref{translatelines}) it is clear that the
phase space point operators can be obtained as a translation of the operator 
associated with the phase space origin as
\begin{equation}
A(\alpha)=T_\alpha A(0)T^\dagger_\alpha.\label{aalphaa0}
\end{equation}
In turn, the phase space point operator $A(0)$ is obtained
as a sum over projection operators onto the states associated with the rays
(lines crossing the origin): 
\begin{equation}
A(0)={1\over N}(\sum_{\lambda/\ 0\in\lambda} P^{(\lambda)}_0 -\mathbb{I}).\label{a00}
\end{equation}
An explicit expression of $A(0)$ as a function of the translation operators can be
obtained by using (\ref{proyector}) in (\ref{a00}). For $n=2$ qubits this operator
can be written as:
\begin{eqnarray}
A(0)&=&\frac{1}{16}\big\{ \sum_{i,j=0}^1(X_0^iX_1^j+Z_0^iZ_1^j)+\nonumber\\
&+& \sum_{m=0}^{2} U_{s}^m
[Y_0+Y_1+Y_0Y_1] U_{s}^{\dagger m}-\mathbb{I}\big\}.\nonumber
\end{eqnarray}

For $n=3$ qubits the form of $A(0)$ is more involved and turns out to be given by
\begin{eqnarray}
&A(0)&=\frac{1}{64}\big\{ \sum_{i,j,k=0}^1(X_0^iX_1^jX_2^k+Z_0^iZ_1^jZ_2^k)+
 \sum_{m=0}^{6} \nonumber\\
&U_{s}^m [& Y_1+(X_1Z_2+Z_1Y_2-Y_1X_2)(\mathbb{I}+Y_0)]U_{s}^{\dagger m}
-\mathbb{I}\big\}.\nonumber
\end{eqnarray}

Finally, it is worth stressing that as the operators $A(\alpha)$ form a complete
basis, one can expand the density matrix in such basis and obtain
\begin{equation}
\rho=N\sum_\alpha W(\alpha) A(\alpha).\label{rhowa}
\end{equation}
Thus, the Wigner function $W(\alpha)$ is nothing but the coefficient of the expansion
of the state $\rho$ in the basis of phase space point operators $A(\alpha)$.  

In the following sections we will display the Wigner function of various
quantum states. However, it is clear that quantum states corresponding to lines
have Wigner functions with simple properties. Thus, using (\ref{aalpha}) 
we can show that the Wigner function of the quantum state $P(\lambda)$ is
equal to $1/N$ over all points in the line $\lambda$ and is equal to 
zero elsewhere. 

Wigner functions are useful to compute expectation values of operators. In particular, 
computing expectation values of translation operators turns out to be particularly
simple. Thus, we can show that 
\beq
{\rm Tr}(\rho T_\beta)= f_\beta \sum_\alpha W(\alpha) (-1)^{\alpha\wedge\beta},\label{meanvalues}
\eeq
where 
\beqa
\alpha\wedge\beta&\equiv& {\bfq}_\alpha\cdot{\bfp}_\beta-{\bfq}_\beta\cdot{\bfp}_\alpha
\nonumber\\
&=&\sum_i q_{\alpha i} p_{\beta i} - q_{\beta i} p_{\alpha i},\label{wedge}
\eeqa
is the analogous of the vector 
product in phase space (that is equal to the area enclosed by the triangle formed by the 
origin and the points $\alpha$ and $\beta$). The function $f_\beta$ depends on the point
$\beta$ and on the quantum net being defined as $f_\beta=N {\rm Tr}(A(0)T_\beta)$. It is 
useful to point out how to compute $f_\beta$: Let us denote the projector onto 
the state associated with the ray that crosses the point $\beta$ as $P_{\lambda_\beta}$. 
This state, which is fixed by the choice of quantum net, is an eigenstate of $T_\beta$ 
with eigenvalue given by $f_\beta$ which, therefore, is always equal to $\pm 1$ and can
be expressed as:
\beq
f_\beta={\rm Tr}(T_\beta\ P_{\lambda_\beta}).
\eeq

There is another identity between Wigner functions that turns out to be useful in some
calculations. For pure states the following identity is valid for 
any translation $T_\alpha$ (or any other operator):
\beq
|{\rm Tr}(\rho T_\alpha)|^2={\rm Tr}(\rho T_\alpha\rho T^\dagger_\alpha).\nonumber
\eeq
This identity, when written in terms of Wigner functions reads
\beq
|\sum_\beta W(\beta) (-1)^{\alpha\wedge\beta}|^2=\sum_\beta W(\beta) W(\beta+\alpha).
\label{twoways}
\eeq
This equation provides a necessary condition for the Wigner function to describe
a pure state. Although the condition is not sufficient, it is useful because of its
simplicity. In some applications this, together with symmetry arguments turns out to 
be enough to determine the value of $W(\alpha)$ in all phase space points. A necessary 
and sufficient condition for the Wigner function to describe a pure state is obtained
by writing the equation $\rho^2=\rho$ in terms of Wigner functions. This implies,
\beq
W(\alpha)=N^2 \sum_{\beta\gamma} \Gamma_{\alpha\beta\gamma} W(\beta) W(\gamma),\nonumber
\eeq
where $\Gamma_{\alpha\beta\gamma}={\rm Tr}(A_\alpha A_\beta A_\gamma)$. Computing these
three--point coefficients is rather involved. Therefore, 
imposing the above necessary and sufficient condition is not a practical way to 
proceed to find constraints on the possible values the Wigner function. 

The discrete Wigner function described in this section has many properties that are
similar to its continuous counterpart. However, there are also some differences that 
are worth pointing out. For the continuous Wigner function (and also for other 
versions of discrete Wigner functions \cite{MPSpra}) the phase space point operators
are both hermitian and unitary operators (up to a normalization). In fact, in such case
the phase space point operator is a displaced reflection operator. For this reason, the
Wigner function at any given phase space point can be measured by using an interesting 
tomographic technique that is described in some detail in \cite{MPSpra} (and was generalized
in \cite{PRS03}). This tomographic
method is a simple application of a more general technique to determine the expectation 
value of a unitary operator. The method does not require to detect all marginal 
distributions (i.e. to perform a complete tomographic determination of the quantum state). 
However, the discrete Wigner function we use in this paper does not have this property. 
Thus, phase space point operators are generally obtained by displacing the operator $A(0)$,
which is not unitary. For this reason, it is not possible to directly measure the 
Wigner function at any phase space point using the tomographic scheme described 
in \cite{MPSpra}. The way to determine the value of $W(\alpha)$ is by using equation 
(\ref{walpha}): $W(\alpha)$ is fixed once we know the probabilities for all states 
associated with the lines that contain the point $\alpha$. 

There is another difference between ordinary Wigner functions and the ones we are
describing here. The relation between phase space point operators $A(\alpha)$ and 
translation operators $T_\beta$ is usually given in terms of a Fourier transform
(see, for example, \cite{MPSpra}). In our case the relation is by means of a 
different transformation, which is related to the Hadamard transform:
\beq
T_\beta=f_\beta\sum_\alpha \ (-1)^{\alpha\wedge\beta}A(\alpha).\label{trelatedtoa}
\eeq
As mentioned above, the factor $f_\beta=N{\rm Tr}(T_\beta A(0))$ depends on the 
quantum net and takes values which are equal to $\pm 1$. Thus, translations and 
phase space point operators relate to each other by means of an Hadamard--like
transform. It is also interesting to notice that the exponent $\alpha\wedge\beta$
can be written in terms of $GF(2^n)$ invariant objects as follows. This exponent
is defined in terms of the binary strings defining position and momentum coordinates
of the phase space points as $\alpha\wedge\beta={\bfq}_\alpha\cdot{\bfp}_\beta-{\bfq}
_\beta\cdot{\bfp}_\alpha$. As we mentioned above, the basis that we use to 
order the momentum axis should be a multiple of the dual of the canonical basis. 
Let this basis be $f_i=\omega^{-k}\tilde e_i$, i.e. the power $k$ indicates what 
multiple of the dual basis $f_i$ is. Then, the above exponent can be shown 
to be identical to 
\beq
\alpha\wedge\beta={\rm Tr}(\omega^{-k}(q_\alpha p_\beta-q_\beta p_\alpha)).\nonumber
\eeq
The right hand side of the above equation is a basis independent expression which 
is entirely written in terms of field operations. 

Finally, it is also worth noticing that equation (\ref{trelatedtoa}) can be inverted and
the phase space point operator can be written in terms of translations as
\beq
A(\alpha)={1\over N^2}\sum_\beta (-1)^{\alpha\wedge\beta} f_\beta \ T_\beta.\label{arelatedtot}
\eeq
All the dependence of the operators $A(\alpha)$ on the quantum net is contained in the 
function $f_\beta$. Thus, this function which, as mentioned above takes values which 
are $\pm 1$, entirely defines the quantum net.

\section{Constructing the Wigner function from the state symmetries} 

Having constructed the phase space representation for systems of qubits a natural question 
arises: For what kind of problems one expects this to be a useful tool? A first attempt to 
answer this question will be presented in this Section. Thus, we expect this tool to be 
of some usefulness when the quantum state and/or the evolution operator have some degree
of symmetry under phase space translations. In this Section we will discuss three specific
examples for which this is indeed the case. First, we will analyze the phase space representation
of some maximally entangled states which are defined precisely as eigenstates of translation
operators. As a first example we will analyze the case of Bell states for a system of $n=2$ 
qubits and show that these states have rather simple Wigner functions. A more interesting 
example is the case of stabilizer error correcting codes \cite{Chuang}. These codes are also 
naturally formulated
in terms of translation operators. Indeed, the code space is defined as the space of common 
eigenstates of $n-1$ commuting operators $S_j$ ($j=1,\ldots,n-1$). These operators, which define
the stabilizer of the code, are nothing but phase space translation operators. Below, we will 
discuss the phase space representation of the simplest such code (the three qubit error correcting
code against phase errors). In this case, as we will see below, the usefulness of the Wigner 
representation turns out to be less obvious. 
Finally, we present the phase space representation of the celebrated `Mean king problem' 
\cite{Vaidman,Aharonov}. This 
problem has a rather appealing solution when formulated in phase space. 

\subsection{Bell states}

\begin{figure}
\includegraphics[width=4cm]{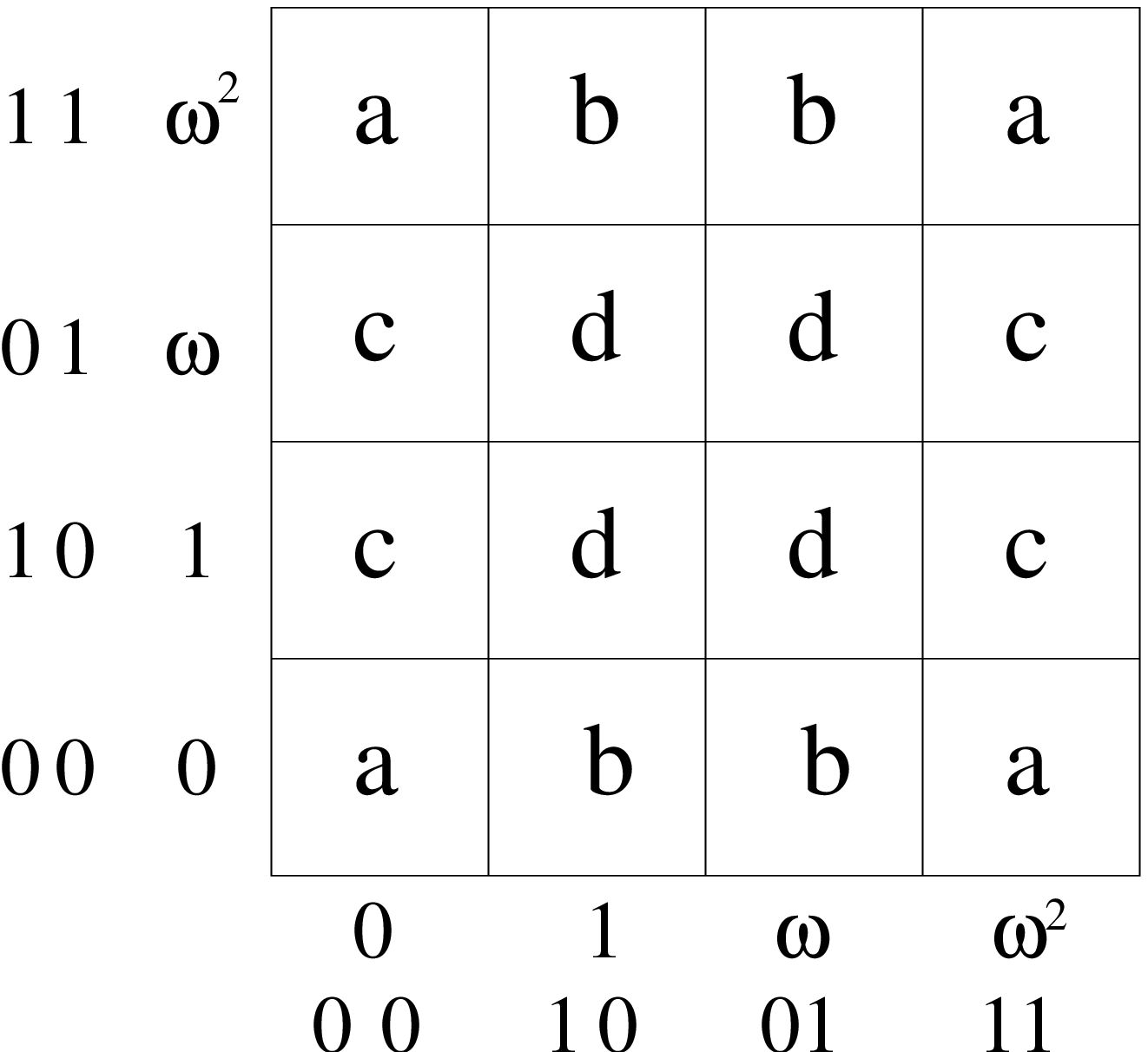}
\caption{Wigner representation for Bell states, depends on four parameters. Using the 
symmetries of the Bell states we reduce our $4 \times 4$ grid of
independent values to the determination of 4 real values $a,b,c$ and $d$.} \label{bsfig1}
\end{figure}

The Bell basis for a system of two qubits is formed by the states
\begin{eqnarray}
|\Psi_\pm\rangle&=&{1\over\sqrt{2}}(|10\rangle\pm|01\rangle),\nonumber\\
|\Phi_\pm\rangle&=&{1\over\sqrt{2}}(|00\rangle\pm|11\rangle).\label{Bell}
\end{eqnarray}
It is simple to show that these states are common eigenstates of the translation operators 
$T_{(\omega^2,0)}=X_0 X_1$ and $T_{(0,\omega^2)}=Z_0 Z_1$ (the eigenvalues of both operators 
are $\pm 1$). For this reason, one expect these
states to have a simple phase space representation. Indeed, the Wigner function of these states
must be invariant under the translations $T_{(\omega^2,0)}$ and $T_{(0,\omega^2)}$. 
The action of these operators is simple:  $T_{( \omega^2,0)}$ interchanges the vertical lines $p=0$ 
and $p=\omega^2$ (and also interchanges the lines $p=1$ and $p=\omega$). 
Similarly, $T_{(0,\omega^2)}$ 
interchanges the corresponding horizontal lines. The total number of points in phase space for two 
qubits is $N^2=16$. Each of the two above symmetries can be use to cut in half the number
of independent parameters that define the Wigner function of a Bell state. Thus, we are left
with only four parameters which characterize the Wigner function as shown in Fig. \ref{bsfig1}. 
It is worth mentioning here that the above results are independent of the quantum net (the 
only assumption we made concerned the association of vertical and horizontal lines with the 
corresponding basis). Indeed, there are $4^3$ quantum networks with Wigner functions having the 
symmetries shown in Fig. \ref{bsfig1}. 

All the four Bell states have Wigner functions with the above symmetries. To find the Wigner
function of each Bell state we need to impose some other conditions that constrain the possible
values of the four parameters $a,b,c,d$. There is an obvious constraint imposed by normalization, i.e.
the sum of all values of the Wigner function must be unity. This implies that 
\begin{equation}
a+b+c+d={1\over 4}.\label{normalization}
\end{equation}
As mentioned above, Bell states are eigenstates of the translation operators with eigenvalues $\pm 1$. 
For the state $|\Phi_+\rangle$ both eigenvalues are $+1$. Imposing this condition is equivalent to requiring 
that the expectation value of the translation operators is equal to $+1$. 
The expectation values of $T_{(0,\omega^2)}$ and $T_{(\omega^2,0)}$ can be easily computed from the 
Wigner function using (\ref{meanvalues}) (in such case, for both translation operators we 
have $f_\beta=+1$). Thus, this condition gives us the following two equations:
\begin{eqnarray}
&a+b-c-d&=\frac{1}{4}, \nonumber \\
&a+c-b-d&=\frac{1}{4}.\label{twoconditions}
\end{eqnarray}
Equations (\ref{normalization}) and (\ref{twoconditions}) are a simple linear system of three 
equations for four unknowns. Therefore, we still have a one parameter family of solutions for the 
Wigner function of each Bell state (the linear system determining the Wigner function for the 
other three Bell states $|\Phi_-\rangle$ and $|\Psi_\pm\rangle$ is formed by (\ref{normalization}) 
and two equations analogous to (\ref{twoconditions}) with the corresponding $\pm$ signs associated
to the eigenvalues). 

\begin{figure}
\includegraphics[width=4cm]{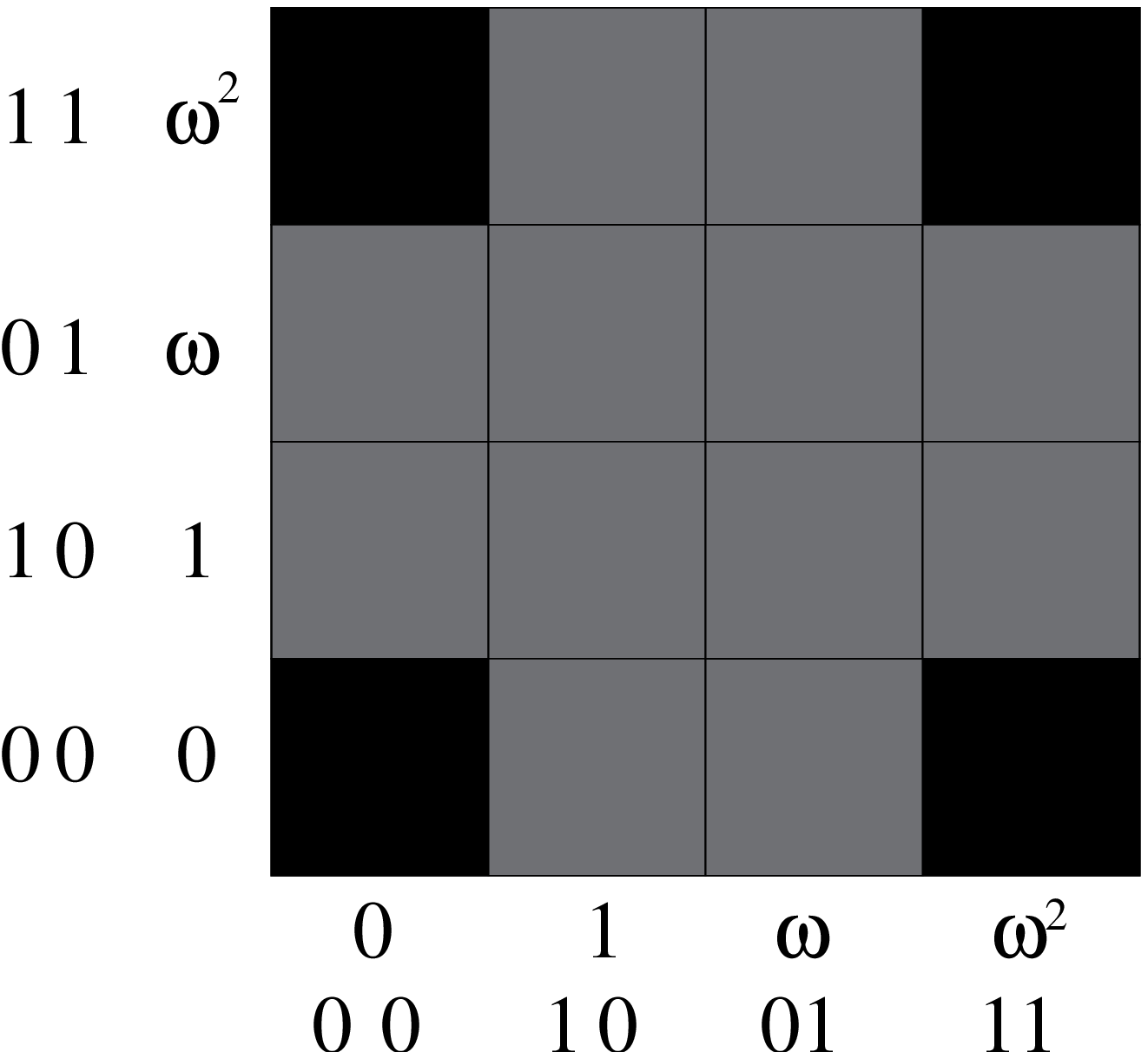}
\includegraphics[height=3.7cm]{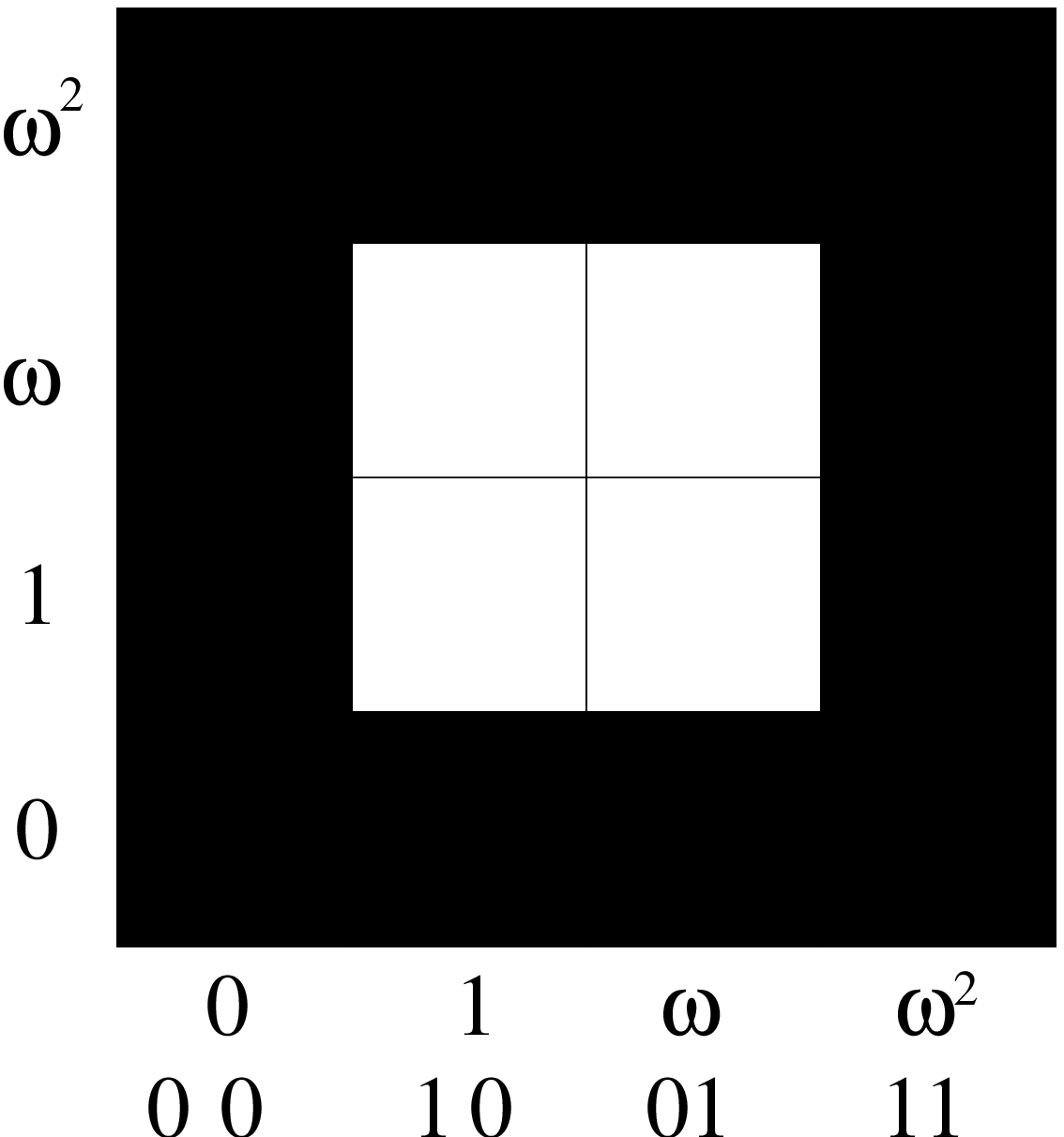}
\caption{The two possible Wigner representations of the Bell state 
$|\Phi_+\rangle=\frac{1}{\sqrt 2}(|00\rangle + |11\rangle)$.
These representations were constructed using the symmetries of the state. The other Bell states can be
constructed by applying the translations $X_1$, $Z_1$ and $X_1Z_1$.} \label{bsfig2}
\end{figure}

It is interesting to notice that the Wigner function of Bell states can be further constrained
by imposing the following condition: Any Bell state is mapped onto an orthogonal state by a unitary
operator that anti-commutes with either $T_{(0,\omega^2)}$ or $T_{(\omega^2,0)}$. This is indeed the 
case for the translations given by $X_0$, $X_1$, $Z_0$, $Z_1$. In fact, applying the translation 
$X_0$ maps $|\Psi_\pm\rangle$ onto $|\Phi_\pm\rangle$. Imposing that the Wigner function of the 
state translated by $X_0$ is orthogonal to the one associated with the original state (i.e., 
that $\sum_\alpha W'(\alpha)W(\alpha)=0$) is equivalent to the following equation:
\begin{equation}
ab+cd=0. \label{orthogonality}
\end{equation}
This equation, together with the above linear set, give two possible solutions: 
$\{a=\frac{1}{4},b=c=d=0\}$ and $\{a=b=c=-d=\frac{1}{8}\}$. The two possible Wigner functions 
are shown in Fig. \ref{bsfig2}. The Wigner functions associated with the other three Bell states
are obtained from the above by applying the corresponding translation operators (for example,
the one for $|\Psi_+\rangle$ is obtained by applying $X_0$, which corresponds to interchanging
the first column with the second one and the third with the fourth). The two solutions presented here
are the only ones allowed by the $4^3$ possible quantum nets. 

\subsection{Phase space representation of a quantum error correction code}

Quantum error correction codes have been under intense investigation during recent years \cite{Chuang}. 
We will not present a detailed introduction to the theory of quantum error correction but simply introduce
the necessary elements to study their phase space representation. 
A wide class of such codes is defined in terms of a {\sl stabilizer} as follows: Let us consider the 
simplest case in which we encode one qubit of quantum information using $n$ physical qubits. 
The space of encoded states is a two--dimensional subspace of the total Hilbert space formed by the set of 
common eigenstates of the operators $S_j$, $j=1,\ldots,n-1$. Such operators are, in the jargon defined
above, a set of commuting translation operators since they are tensor products of Pauli 
operators acting on each qubit. The stabilizer is chosen in such a way that the code corrects 
a set of errors $E_i$ which are also translation operators. The code will correct against errors
$E_i$, $i=1,\ldots,2^{n-1}-1$ if the encoded states $|\phi_L\rangle$ are mapped by the errors 
$E_i$ onto subspaces which are mutually orthogonal for different values of $i$. 

The use of phase space representation in this context seems to be natural. Indeed, as encoded states
are eigenstates of translation operators their corresponding Wigner functions must be invariant under
the same translations. Moreover, as errors are also translation operators, when translating the Wigner
function of an encoded state by a correctable error one should obtain a Wigner function which is 
orthogonal to the original one. Therefore, one expects (perhaps naively) the phase space method 
discussed above to provide some insight into error correction. Also, the task of 
finding the Wigner function of encoded states could be accomplished using the same ideas we 
described above to determine the Wigner function of Bell states (i.e., using the symmetries of 
the state in the first place). 

Here we will discuss a specific example, the simplest stabilizer quantum error correcting code. The 
code encodes one qubit of quantum information using $n=3$ physical qubits and corrects against 
errors of the form $Z_0$, $Z_1$ and $Z_2$ (i.e., phase errors). The stabilizer of the code is 
defined by the translation operators 
\begin{eqnarray}
S_1&=&T_{(\omega^6,0)}=X_0 X_1,\nonumber\\
S_2&=&T_{(\omega^5,0)}=X_1 X_2.\label{stabilizer}
\end{eqnarray}
The reason why the code can correct against all $Z_i$ errors is that each error maps eigenstates
of $S_i$ onto eigenstates of the same operators with different eigenvalues. Indeed, if we define
the code space as the set of all states with eigenvalues $+1$ for both $S_1$ and $S_2$, the action 
of errors on encoded states turns out to be the following: $Z_0$ errors maps encoded states onto 
eigenstates of $S_1$ and $S_2$ with eigenvalues $-1$ and $+1$ while for $Z_1$ and $Z_2$ errors the 
corresponding eigenvalues are $s_1=-1$, $s_2=-1$ and $s_1=+1$, $s_2=-1$ respectively. 

\begin{figure}
\includegraphics[width=6cm]{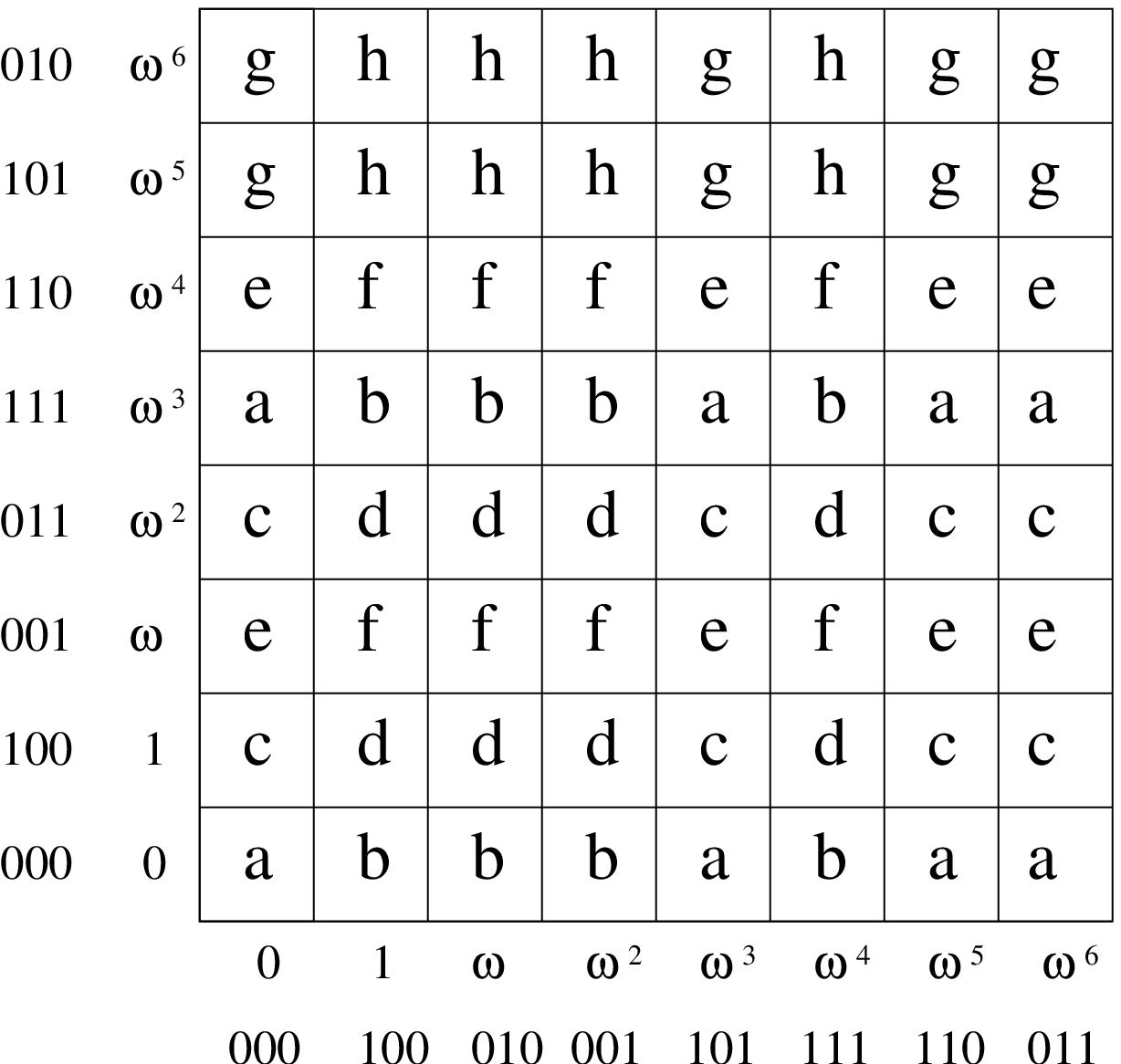}
\caption{Wigner function for a logical state. The number of independent variables in 
this function can be reduced by using the symmetries of the state under the stabilizer
translations. In the $8 \times 8$ grid there are only $8$ independent variables to 
be determined ($a,b,c,d,e,f$ and $g$).} \label{ecfig1}
\end{figure}

As mentioned above, the Wigner function of encoded states must be symmetric under the translations
$S_1$ and $S_2$. As was the case for Bell states, each of these symmetries cuts in half the total 
number of independent parameters defining the Wigner function. So, invariance under $S_1$ and $S_2$ 
is achieved only if the Wigner function is identical along the following vertical lines:
$W(0,p)=W(\omega^3,p)=W(\omega^5,p)=W(\omega^6,p)$, $W(1,p)=W(\omega,p)=W(\omega^2,p)=W(\omega^4,p)$
(the identities must hold for all values of $p$). Thus, this symmetry reduces the number of 
parameters defining the Wigner function of encoded states from $8\times 8$ to $8\times 2$. 
Below, we will display the Wigner function of general encoded states. But before that, it is 
simpler to start by analyzing the Wigner function of specific encoded states. To define the states 
encoding the two logical states $|0_L\rangle$ and $|1_L\rangle$ we can proceed as follows: These 
states can be chosen to be encoded states (which are eigenstates of eigenvalue $+1$ of $S_1$ and 
$S_2$) which are also eigenvalues of a third operator that commutes with the stabilizer. This 
operator can be chosen in this case as $Z_L=Z_0Z_1Z_2$, which is also a translation operator. 
Imposing invariance under $Z_L$ again cuts in half the number of independent parameters. Thus, as 
$Z_L$ interchanges horizontal lines one must identify the values of the Wigner function along such 
lines (the $Z_L$ symmetry implies that $W(q,0)=W(q,\omega^3)$, $W(q,1)=W(q,\omega^2)$, 
$W(q,\omega)=W(q,\omega^4)$, $W(q,\omega^5)=W(q,\omega^6)$). Therefore, the Wigner function of the 
two logical states can be parametrized with $8$ parameters and has the symmetries displayed in 
Fig. \ref{ecfig1}. 

We can apply the same reasoning we used to find the Wigner function of Bell states above to find extra 
conditions on the above eight parameters. First, we can impose the normalization condition and the 
fact that logical states have eigenvalue $+1$ of the translation operators $S_1$ and $S_2$. For the 
case of the state $|0_L\rangle$, the eigenvalue of $Z_L$ is also equal to $+1$ and the 
corresponding conditions turn out to be
\begin{eqnarray}
a+b+c+d+e+f+g+h&=&\frac{1}{8}, \nonumber \\
a+b+c+d-e-f-g-h&=&\frac{1}{8}, \nonumber \\
a+b-c-d+e+f-g-h&=&\frac{1}{8}, \nonumber \\
a-b+c-d+e-f+g-h&=&\frac{1}{8}. 
\end{eqnarray}
The conditions defining the logical state $|1\rangle_L$ are the same as above except 
for the last equation 
where the right hand side is $-1/8$. It is worth noticing that once we have the Wigner 
function of the 
logical state $|0_L\rangle$ the one for $|1_L\rangle$ is obtained by translating it 
with the operator $X_0$ which interchanges vertical lines. 

\begin{figure}
\includegraphics[width=6cm]{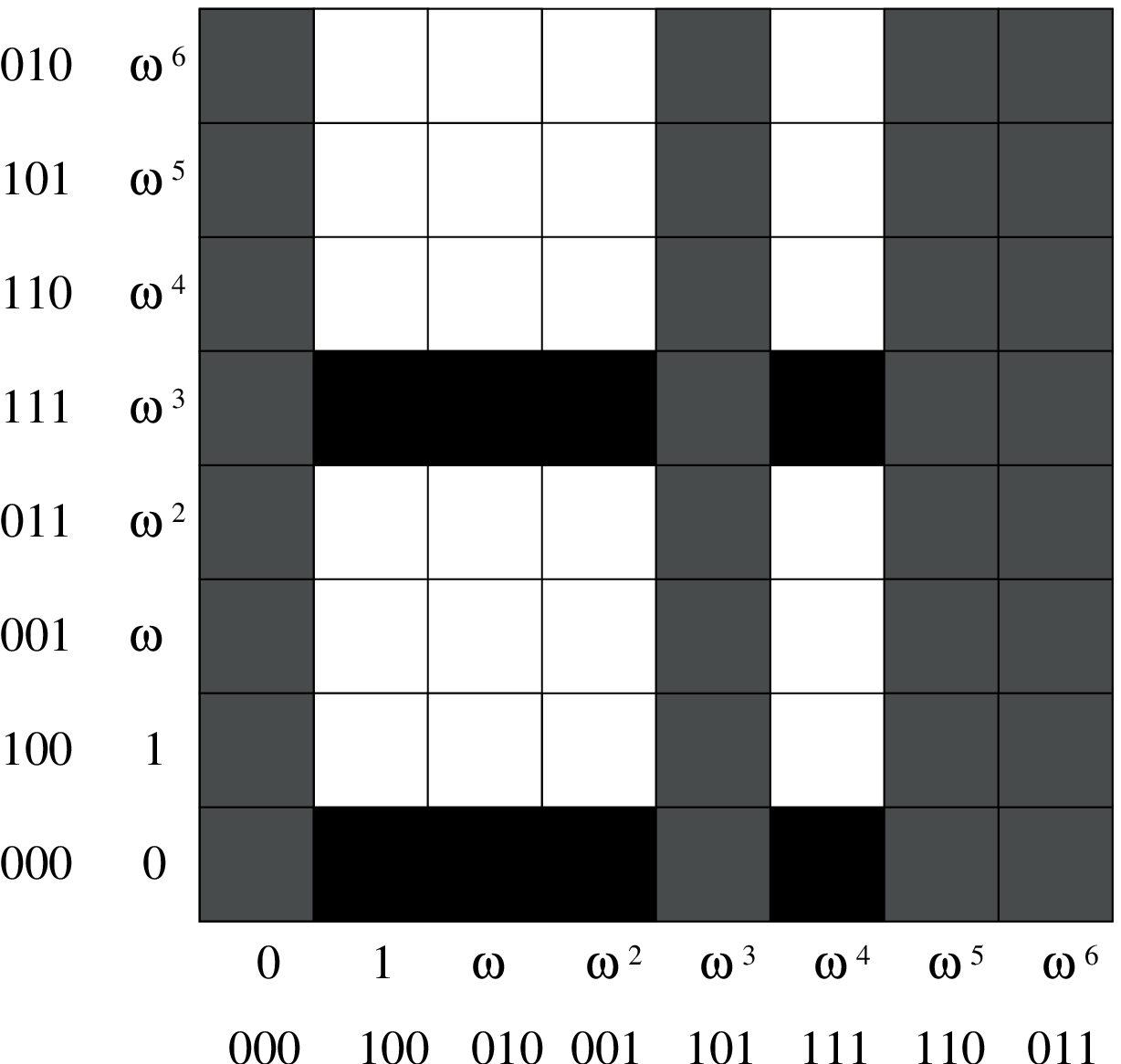}
\includegraphics[width=6cm]{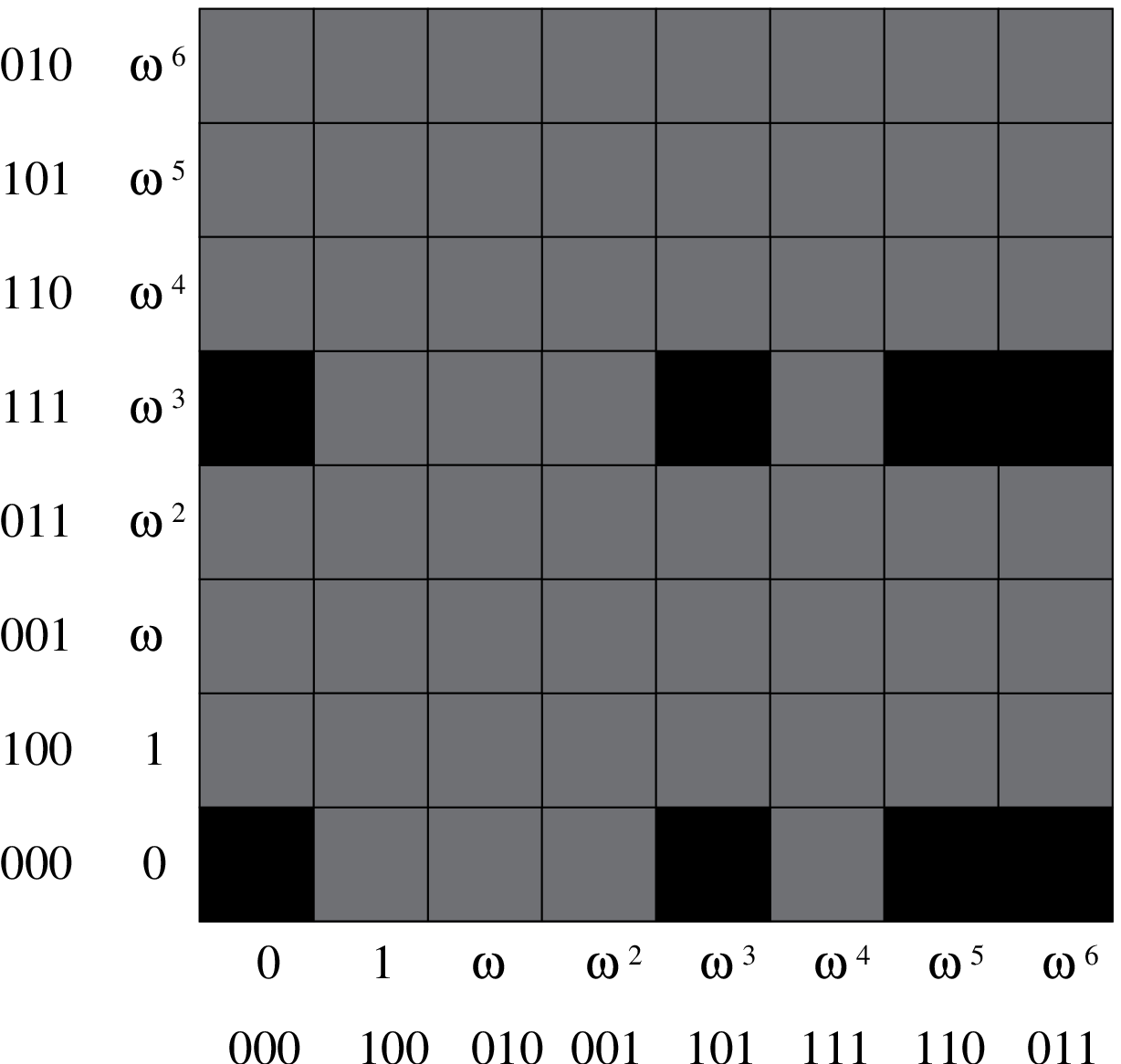}
\caption{Two possible Wigner representations of the state $|0\rangle_L$. The one at the top, 
where $\{a=c=e=g=\frac{1}{32}\}$, is covariant under the action of $U_\omega$. The quantum
net is defined by associating the eigenstate $Z_1|\lambda_0\rangle$ to the main diagonal 
line (where $|\lambda_0 \rangle$ is the eigenstate with
eigenvalue +1 of the generators). The one at the bottom, where 
$\{a=\frac{1}{8}, e=c=g=0\}$, is not covariant under $U_\omega$. The color 
convention is such that black (white) regions correspond to positive (negative) values of the
Wigner function.} \label{ecfig2}
\end{figure}

The above linear system still allows for a four parameter family of solutions. Extra 
conditions can be
imposed in the same way as we did for Bell states. Indeed, errors 
$Z_0$, $Z_1$ and $Z_2$ are such that
they transform encoded states into orthogonal states. Imposing the orthogonality with the 
translated Wigner functions yields the following set of equations
\begin{eqnarray}
ae+bf+cg+dh&=&0, \nonumber \\
ac+bd+eg+fh&=&0, \nonumber \\
ag+bh+ce+df&=&0.
\end{eqnarray}
Finally, to find solutions that correspond to quantum states we should impose the condition that 
the sum of values of Wigner function along arbitrary lines should always be non--negative. Also, 
we impose the condition for the state to be pure (i.e., $N\sum_\alpha W^2(\alpha)=1$), which 
is equivalent to 
\begin{equation}
a^2+b^2+c^2+d^2+e^2+f^2+g^2+h^2=\frac{1}{64}.
\end{equation}
After some algebra we find that the possible solutions must obey the following conditions: 
\begin{equation}
b=\frac{1}{8}-a,\  f=-e,\  d=-c,\  h=-g.\label{conditions}
\end{equation} 
We find eight solutions with all the desired properties. They are
\begin{eqnarray}
a&=&\frac{1}{8}, e=c=g=0;\nonumber\\
a&=&e=\frac{1}{16}, c=g=0;\nonumber\\
a&=&c=\frac{1}{16}, e=g=0;\nonumber\\
a&=&g=\frac{1}{16}, c=e=0;\nonumber\\
a&=&c=e=g=\frac{1}{32};\nonumber\\
a&=&\frac{3}{32}, c=e=-g=\frac{1}{32};\nonumber\\
a&=&\frac{3}{32}, c=-e=g=\frac{1}{32};\nonumber\\
a&=&\frac{3}{32}, -c=e=g=\frac{1}{32}.\label{solutions}
\end{eqnarray}
So far we did not impose any condition on the quantum net. In fact, some of the Wigner functions obtained
in this way do not have the property of covariance under the operation $U_\omega$. Imposing this 
condition we are left with only four solutions which are the ones corresponding to the 
last four equations. In Fig. \ref{ecfig2} we show one of these solutions, and the solution
where the Wigner function only takes positive values. 

\begin{figure}
\includegraphics[width=3cm]{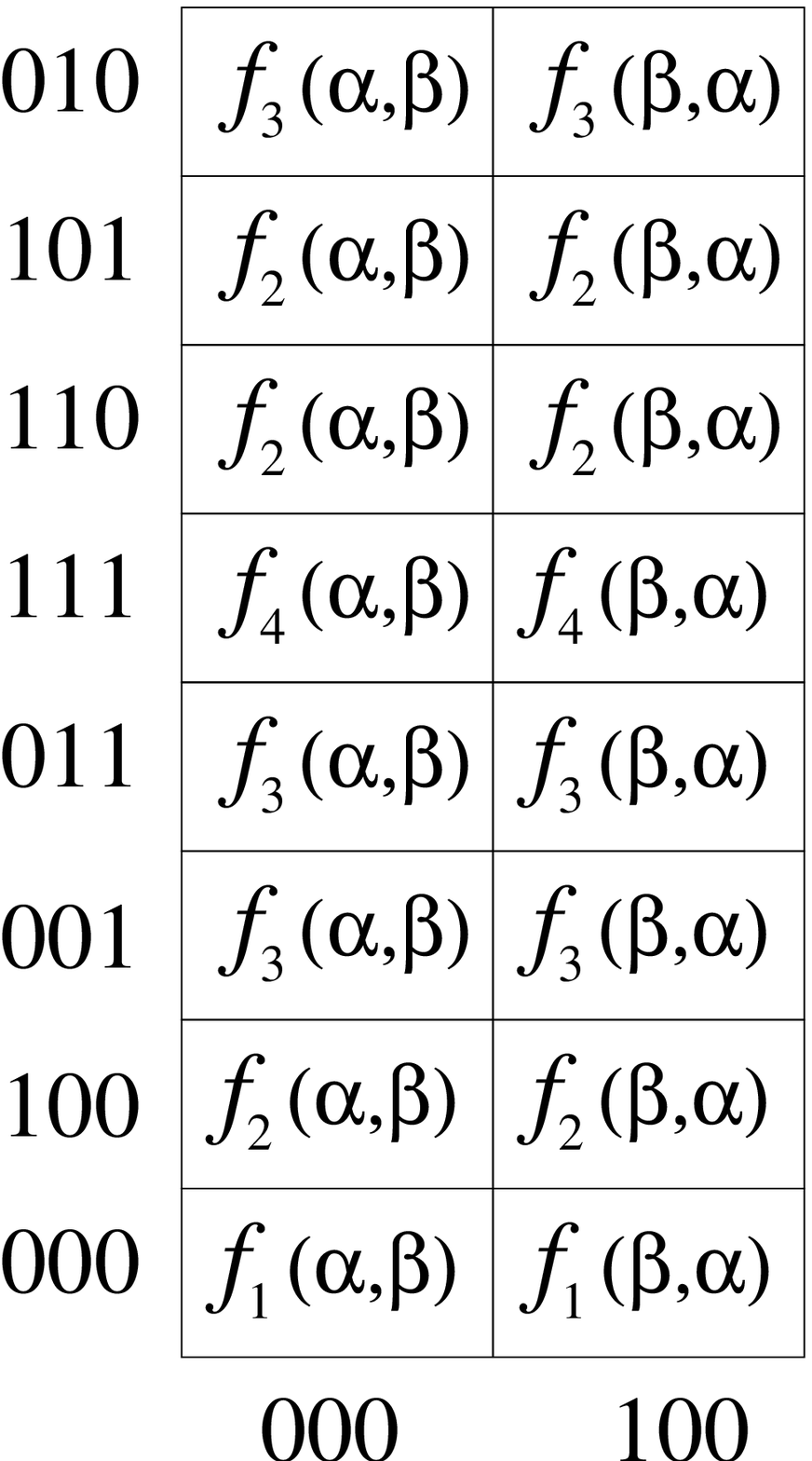}
\caption{The value of the Wigner function of a general encoded 
state $\alpha |0\rangle_L+\beta |1\rangle_L$ in the first two columns of the phase 
space. The quantum net is the same one used
at the top of Fig. \ref{ecfig2}. The other columns can be obtained from the invariance
of the state under the stabilizer translations $S_1$ and $S_2$.} \label{generallogic}
\end{figure}

Finally, we completely specified the quantum net by choosing the state corresponding to the main 
diagonal to be $Z_1|\lambda_0\rangle$, where $|\lambda_0 \rangle$ is the eigenstate with
eigenvalue $+1$ of the three generators corresponding to this line. For this quantum net 
the Wigner function of the encoded state $|0\rangle_L$ is the one shown at the top of Fig. \ref{ecfig2}.
For this case, we also obtained the Wigner function of the most general encoded state 
$|\phi_L\rangle=\alpha |0_L\rangle+\beta |1_L\rangle$. This Wigner function 
is displayed in Fig. \ref{generallogic} and is completely determined by the following 
four functions:
\begin{eqnarray}
f_1(\alpha,\beta)&=&\frac{1}{32}[|\alpha|^2+3|\beta|^2+(2+i)\alpha \beta^* +(2-i) \alpha^* \beta], \nonumber \\
f_2(\alpha,\beta)&=&\frac{1}{32}[|\alpha|^2-|\beta|^2+i(\alpha \beta^* - \alpha^* \beta)], \nonumber \\
f_3(\alpha,\beta)&=&\frac{1}{32}[|\alpha|^2-|\beta|^2-i(\alpha \beta^* - \alpha^* \beta)], \nonumber \\
f_4(\alpha,\beta)&=&\frac{1}{32}[|\alpha|^2+3|\beta|^2-(2+i)\alpha \beta^* -(2-i) \alpha^* \beta].\nonumber
\end{eqnarray}

The Wigner function shown in Fig. \ref{generallogic}
has all the symmetries defining encoded states
(i.e., it is invariant under interchange of vertical lines corresponding to the translation operators
$S_1$ and $S_2$ and is mapped onto an orthogonal state when translated by 
errors $Z_0$, $Z_1$ and $Z_2$). 
Our naive expectation was that these properties were going to be more evident in the solution we
obtained. However, this is not the case, which casts doubts about the usefulness of the phase space
representation for quantum error correction.

\subsection{The Mean King Problem: a phase space solution}

Here we present a third application of phase space tools to quantum information. 
We will solve the so--called {\sl mean king problem}, that was first presented in 
\cite{Vaidman}. The formulation of the problem is the following: A physicist must 
prepare a spin $1/2$ particle in a state of his choice. Then he should give the particle
to the {\sl mean king}. The king makes a projective measurement of one of the 
three Cartesian components of the spin of the particle (i.e., the king measures either 
the $X$, the $Y$ or the $Z$ observable). Then, the king gives the particle back 
to the physicist who is allowed to perform any operation on it. Finally, the king announces
what observable was measured in his laboratory. The physicist would only save
his life if he is able to retrodict the result of the king measurement based on the results
of the measurements performed in his own laboratory before knowing the observable measured
by the king. This problem was also extended to cases where the physicist is given  
quantum systems with a Hilbert space with dimension which is a power of a 
prime \cite{Aharonov}. The solution of the mean king problem is only possible 
if the physicist entangles the particle to be sent to the king with another identical 
particle he keeps in his own laboratory. 

\begin{figure}[ht]
\includegraphics[width=8cm]{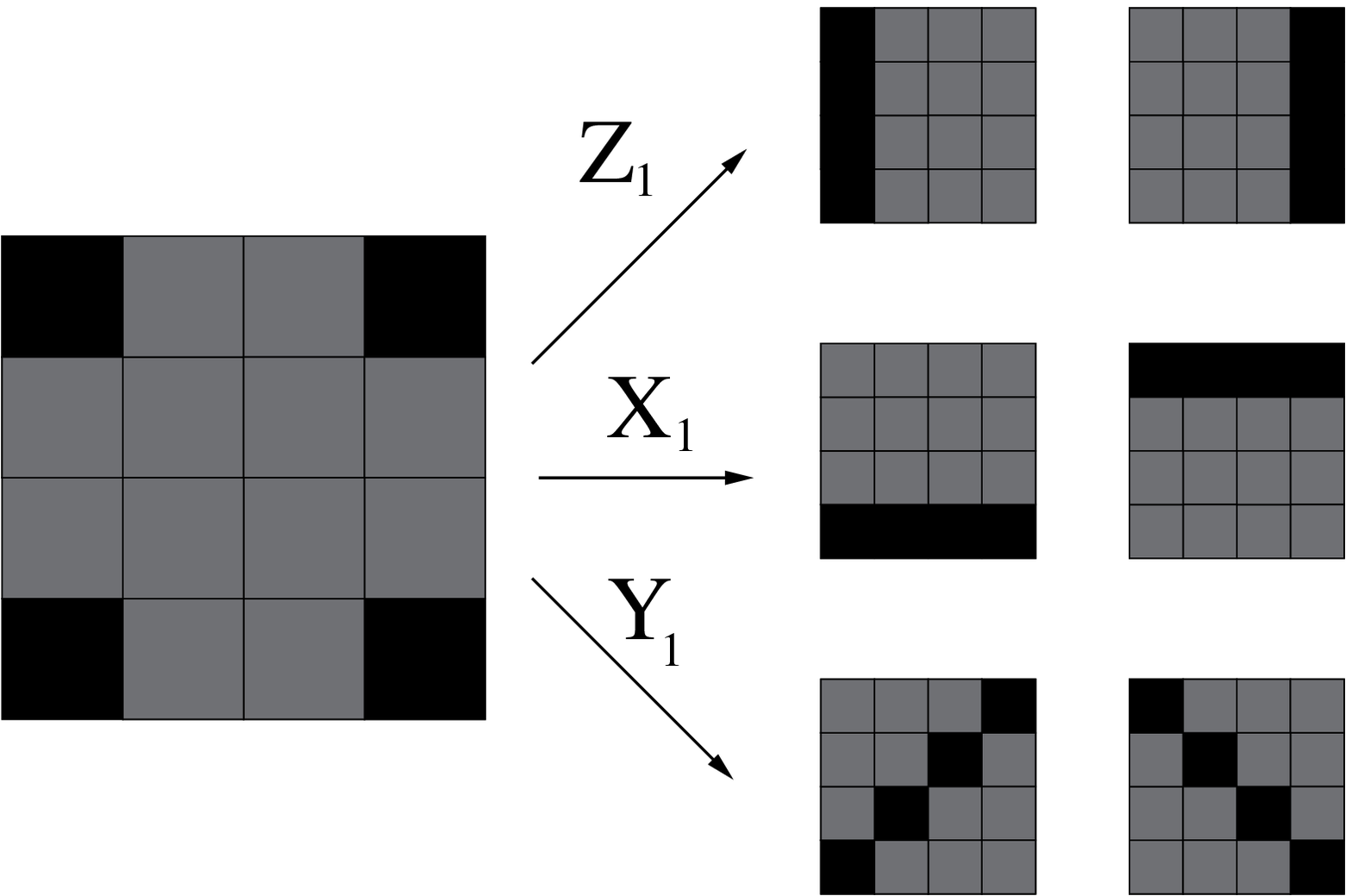}
\caption{A Wigner representation of the state $|\Phi_+\rangle$. At the right we show 
the possible states that the king could give us according to his measurement. The 
six states are associated to the vertical lines $v_1$ and $v_2$, to the horizontal
lines $h_1$ and $h_2$ or to the diagonal lines $d_1$ and $d_2$.}\label{mkfig1}
\end{figure}

We will analyze this problem using phase space methods, which seem to be well suited 
for this purpose. Indeed, maximally entangled (Bell) states are naturally 
represented in phase space as discussed 
above. Moreover, the observables measured by the king are also naturally represented in 
phase space since they are translation operators. For this reason, one may suspect that 
phase space methods may enable a simple solution to the problem. This is indeed the case, 
as we will now discuss. 
We will not present here the usual solution to the problem but will attempt to 
present a solution entirely based on phase space. Let us consider
the initial state prepared by the physicist to be the Bell state $|\Phi_+\rangle$ whose
Wigner function was displayed above. When the king measures one of the three components
of the spin of the first particle he can obtain one of two values ($\pm 1/2$). Each of 
these measurements can be viewed as the projection of the original state onto a basis of
``line states'': Vertical lines are associated with the measurement of $Z$, horizontal lines
are associated with the measurement of $X$ and the striation containing the ray corresponding
to the main diagonal of phase space corresponds to the measurement of $Y$. From the 
Wigner function of the above Bell state it is clear that the initial state has non-vanishing
projection only on two states of each of these three striations. Therefore, as a 
result of his measurement, the king prepares one of six states which are displayed 
in Fig. \ref{mkfig1}. The quantum net we use here is such that the states corresponding 
to the two vertical lines are $|11\rangle_z$ (right vertical line denoted as $v_1$)
and $|00\rangle_z$ (left vertical line, denoted from now on as $v_2$), the two 
states corresponding to the horizontal lines are $|11\rangle_x$ (top horizontal line: $h_1$)
and  $|00\rangle_x$ (bottom horizontal line, denoted as $h_2$) 
and the two states corresponding to the diagonal lines are $|10\rangle_y$ 
(diagonal not crossing the origin: $d_1$) and $|01\rangle_y$ (diagonal crossing
the origin, denoted as $d_2$). 

Once the king makes his measurement and gives the particle back to the physicist, he should 
device a measurement scheme which would enable him to retrodict the result of the king's 
measurement once he is informed about the measured observable. What can the physicist do? The 
solution is to measure a collective observable of the two particle system with four distinct 
eigenvalues. Each of the four eigenstates of this observable must have vanishing 
overlap with one and only one of the two states generated by each measurement. 
In such case, when the king announces what was the measured observable the 
physicist will always be able to infer what was the measured result (since there is only 
one result of every king measurement which would be consistent with the result of 
his own measurement). For example, one such state, 
which we denote as $|\varphi_1\rangle$,  should be orthogonal to the line 
states associated with the upper horizontal line ($h_1$), to the rightmost vertical line 
($v_1$) and to the diagonal line not crossing the origin ($d_1$). 
As mentioned above these three line states are $|11\rangle_z$, $|11\rangle_x$ 
and $|10\rangle_y$. Below, we will show how to obtain the Wigner function of this state. 
It is important to notice that once we obtain this function we can find the 
Wigner functions associated to the other three states that complete the basis of 
the Hilbert space by implementing simple translations. This is the case for the 
following reason: The six line states 
relevant for our construction are connected by phase space translations. For example, 
the operator $T_{(\omega^2,0)}=X_0X_1$ interchanges the two vertical states and the two 
diagonal states. Similarly, the operator $T_{(0,\omega^2)}=Z_0Z_1$ interchanges the two 
horizontal states and the two diagonal states while $T_{(\omega^2,\omega^2)}$ interchanges
the two vertical states and the two horizontal ones. Therefore, it is natural to 
require that $|\varphi_1\rangle$ and its three translated descendants form an orthogonal 
basis of the Hilbert space. This basis defines the observable to be measured by the 
physicist.

\begin{figure}[ht]
\includegraphics[width=4cm]{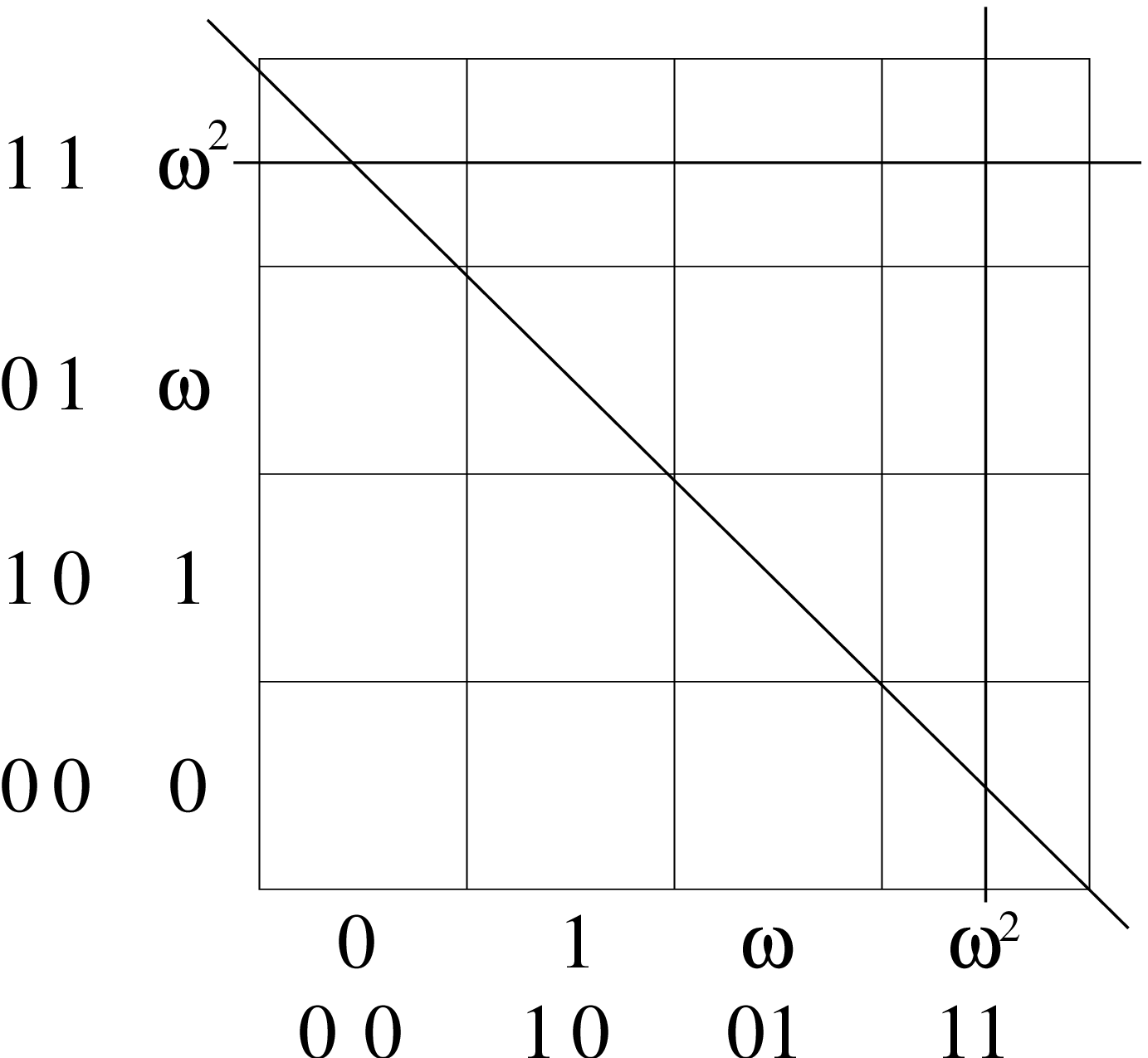}
\caption{The Wigner function of the state $|\varphi_1\rangle$ that defines the solution 
to the mean king problem should add up to zero along the lines indicated in the 
Figure. As it is evident from the graph, 
is natural to impose that the Wigner function is invariant under reflection 
about the main diagonal.}\label{king-lines}
\end{figure}

To find the Wigner function of the state $|\varphi_1\rangle$ we need to determine its value
in all the $4\times 4$ points of the phase space grid. For this purpose, we can 
proceed much in the same way we did in the previous sub--sections. Thus, we 
impose the following conditions: The first condition tells us that the sum of 
the values of the Wigner function along the three lines $h_1$, $v_1$ and $d_1$ must be 
equal to zero. Fig. \ref{king-lines} shows the three lines that define this states. 
This condition imply the following set of three linear equations 
\begin{equation}
\sum_{\alpha\in h_1}W(\alpha)=\sum_{\alpha\in v_1}W(\alpha)=\sum_{\alpha\in d_1}W(\alpha)=0.
\end{equation}
The second condition arises by imposing that the three translation operators 
$T_{(\omega^2,0)}$, $T_{(0,\omega^2)}$ and $T_{(\omega^2,\omega^2)}$ map the state into 
an orthogonal one. This is equivalent to impose that the expectation value of these
operators is equal to zero. In turn, this condition is naturally expressed in terms 
of the Wigner function as:
\begin{eqnarray}
\langle X_1X_2\rangle=0&\iff&
 \sum_{\alpha\in h_1,h_2}W(\alpha)=\sum_{\alpha\notin h_1,h_2}W(\alpha),\nonumber\\
\langle Z_1Z_2\rangle=0&\iff&
 \sum_{\alpha\in v_1,v_2}W(\alpha)=\sum_{\alpha\notin v_1,v_2}W(\alpha),\nonumber\\
\langle Y_1Y_2\rangle=0&\iff&
 \sum_{\alpha\in d_1,d_2}W(\alpha)=\sum_{\alpha\notin d_1,d_2}W(\alpha).\nonumber
\end{eqnarray}
It is simple to show that these conditions, together with the normalization condition, 
imply that the sum of the Wigner function along the lines $h_2$, $v_2$ and $d_2$
must be equal to $1/2$, i.e. 
\begin{equation}
\sum_{\alpha\in h_2}W(\alpha)=\sum_{\alpha\in v_2}W(\alpha)=
\sum_{\alpha\in d_2}W(\alpha)={1\over 2}.
\end{equation}
Also, for the same reason the sum of the Wigner function along the remaining two lines
of each striation (i.e., $h_{3,4}$, $v_{3,4}$ and $d_{3,4}$) 
must be equal to $1/2$, i.e.
\begin{equation}
\sum_{\alpha\in h_3, h_4}W(\alpha)=\sum_{\alpha\in v_3, v_4}W(\alpha)=
\sum_{\alpha\in d_3, d_4}W(\alpha)={1\over 2}.
\label{c4}
\end{equation}
Clearly, the above conditions are not enough to completely determine the Wigner function. 
However, we can obtain a solution by noticing that the state $|\varphi_1\rangle$ has an 
extra symmetry. In fact, it is invariant under the operation $P_{12}H_1H_2$ which 
interchanges the two particles after applying a Hadamard transformation to each 
of them. This is so because the Hadamard operation applied to both qubits 
interchanges lines $h_1$ with $v_1$ and $d_1$ with $d_2$. On the other hand, the 
permutation leaves $h_1$ and $v_1$ invariant while interchanging $d_1$ and $d_2$. 
Using this symmetry, together with the previous results, we can show that the sum 
of the Wigner function along lines $h_3$ and $h_4$ has the same value (similarly for 
the remaining vertical and diagonal lines). 

\begin{figure}[ht]
\includegraphics[width=3cm]{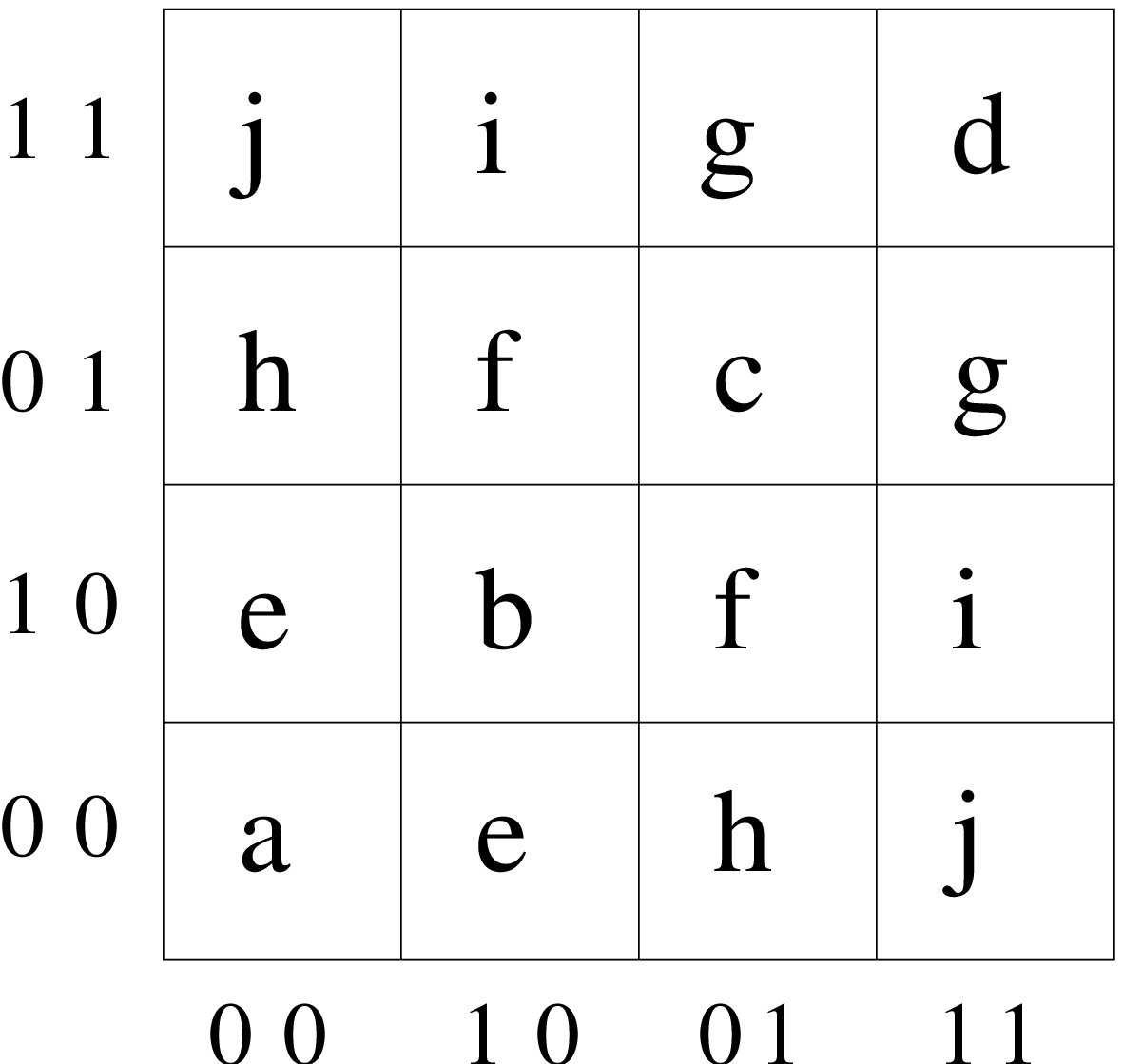}
\caption{Ansatz for the Wigner function of the state $|\varphi_1\rangle$ defining the
solution of the mean king problem. It depends upon $10$ real parameters and is symmetric
upon reflection about the main diagonal.}\label{mkfig2}
\end{figure}

\begin{figure}[ht]
\includegraphics[width=4cm]{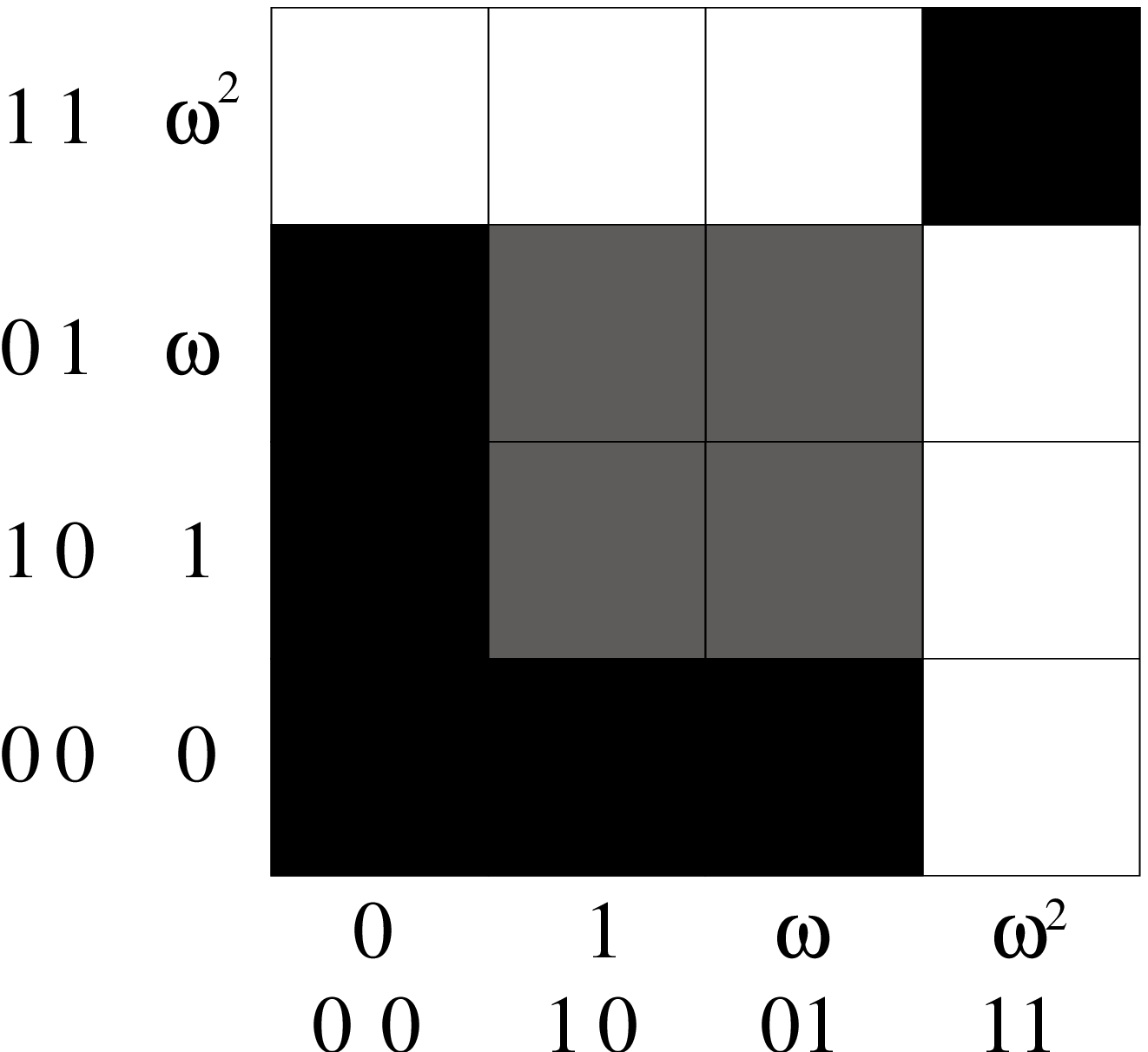}
\caption{The Wigner function of the state $|\varphi_1\rangle$, which defines the solution 
to the Mean King Problem. The physicist needs to measure an observable which is diagonal
in the basis formed by this state and the ones obtained by translating it using
the operators $Z_0Z_1$, $Y_0Y_1$ and $X_0X_1$. The color convention 
is such that positive (negative) values of the Wigner function correspond to 
black (white) regions. The solution is such 
that $a=e=d=h=3/16$, $b=c=f=1/6$, $j=i=g=-1/16$.}\label{mkfig3}
\end{figure}

So, all the conditions discussed so far can be summarized as follows: In each of the three
striations (vertical, horizontal and ``diagonal'') the Wigner function adds up to $1/2$ in 
one line, it adds up to $1/4$ in two other lines and it adds up to $0$ in the remaining line. 
It is simpler to find a solution that has the same symmetry shown in Fig. 
\ref{king-lines}, \textit{i.e.} reflection about the main diagonal of phase space. For this 
reason, the Wigner function we will obtain can be parametrized by ten real numbers
as shown in Fig. \ref{mkfig2}. The ten parameters should obey all the conditions
listed above, which end up giving rise to seven independent linear equations. To 
fix the solution we need to impose other constraints. The need for these constraints
comes from the fact that so far we did not impose the Wigner function to describe a
state. This can be done by imposing, for example, that the purity is equal to one 
(i.e., $N\sum_\alpha W^2(\alpha) =1$). Also, we can use the fact that the orthogonality
between $|\varphi_1\rangle$ and the translated state $|\varphi_2\rangle=
T_{(0,\omega^2)}|\varphi_1\rangle$ implies that $\sum_\alpha W_1(\alpha)W_2(\alpha)=0$. 
It is worth stressing that this condition is equivalent to ${\rm Tr}(\rho T_{(0,\omega^2)})=0$
only if the state is pure. Finally, we can impose that the Wigner function can be
used to compute the expectation value of, for example, the operator $X_0$ in two 
equivalent ways, i.e.
\beq
|\sum_\beta W(\beta) (-1)^{\alpha\wedge\beta}|^2=N\sum_\beta W(\beta)
W(\beta+\gamma).\nonumber
\eeq
where $\gamma=(1,0)$. The above set of conditions give four possible consistent 
Wigner functions. The one corresponding to the quantum net which is covariant 
under the operator $U_\omega$ is given in Fig. \ref{mkfig3}.

\section{Conclusions}

In this paper we reviewed the discrete Wigner function for
systems of $n$ qubits. We showed a rather simple way to construct the phase space for
these systems. The matrix representation of the finite field $GF(2^n)$ and the 
quantum circuit representation of unitary operators are two very useful tools in this 
context. The position and momentum axis of phase space are labeled with elements
of the field $GF(2^n)$. To associate each vertical line with a computational state
one associates each field element with the $n$--tuple formed by the coordinates of the
field element in a given basis. The same criterion is applied for the horizontal lines. 
In this approach there is a unitary operator $U_\omega$ that plays a crucial role: $U_\omega$ is
the unitary operator that permutes computational states in such a way that while fixing
the state associated with the first vertical line ($q=0$) it maps every other 
computational state onto the one located immediately ``to the right''. The same operator
maps momentum states moving them ``downwards'' and therefore corresponds to the 
classical squeezing operation that maps a point $(q,p)$ onto another point 
$(\omega q,p/ \omega)$. 
It turns out that not only this operator can be easily represented in terms of a 
simple circuit entirely made out of control--not and swap gates. Also, the operator 
provides the change of basis between the $N-1$ mutually unbiased bases which are 
associated with all the (non--vertical or non--horizontal) striations. In fact, 
$U_\omega$ can be used to explicitly generate all the states of $N-2$ bases given the 
states of a single one (for example, defining the states associated with the striation
corresponding to the main diagonal of phase space one would obtain the remaining ones
applying powers of $U_\omega$). It is important to stress that imposing covariance of 
the Wigner function under this kind of ``squeezing'' transformation  
reduces substantially the ambiguity in defining the quantum net (but does not
completely fixes it).   

We also presented here some new applications of the phase space formalism. 
We studied the phase space representation of quantum stabilizer codes (or states). 
As these states are eigenstates of a set of translation operators their Wigner 
function exhibit some obvious symmetries. The approach described in the paper, that
enabled to find the Wigner function from the state symmetries, can be generalized 
in some obvious ways. Thus, we can show that for a stabilizer state the symmetry 
of the state under phase space translations implies that the Wigner function 
in the entire $N\times N$ grid can be constructed from the value it takes 
in $N$ points. Moreover, the value of the Wigner function for stabilizer states 
is always an integer multiple of $1/N^2$. Thus, for stabilizer states we can 
obtain a simple formula for the Wigner function as follows: A stabilizer state 
is defined to be a pure eigenstate of a set of $N$ commuting translation operators. 
Consider the state $\rho_S$ to be such that ${\rm Tr}(\rho_S T_\beta)=g_\beta=\pm 1$ 
for $\beta\in S$. That is to say, the state is an eigenstate of $T_\beta$ with 
eigenvalue given by $g_\beta=\pm 1$ if the index belongs to a set $S$. For all 
other translation operators the same state has vanishing expectation value, i.e. 
${\rm Tr}(\rho_S T_\beta)=0$ if $\beta\notin S$. 
Therefore, using equation (\ref{arelatedtot}) the Wigner function of a stabilizer state is
\beq
W_S(\alpha)={1\over N^2} \sum_{\beta\in S} f_\beta g_\beta (-1)^{\alpha\wedge\beta}.
\label{wignerstabilizer}
\eeq
This result is valid for all stabilizer states and for all phase space points. From this
equation it is evident that the Wigner function has the same value for all the points of 
the stabilizer: 
Thus, if we consider $\alpha\in S$ we must use $\alpha\wedge\beta=0$ in (\ref{wignerstabilizer})
since the corresponding translations commute. 
In such case we have $W_S(\alpha\in\ S)=\sum_{\beta\in S}
f_\beta g_\beta/N^2$. Moreover, we can show that for every stabilizer $S$ the quantum 
net can be chosen in such a way that $f_\beta=g_\beta$ (for $\beta\in S$). Then, 
the Wigner function is equal to $1/N$ in the points of the stabilizer and is equal to 
zero everywhere else. These features are evident in the examples we discussed above where 
we analyzed Bell states
and encoded states of the three qubit error correcting code.  However, the results we 
obtained are not conclusive about the potential usefulness of the phase space 
approach as a natural tool for this problem (even from a purely pictorial point of view). 
More work along this line is in progress. The phase space solution of the 
``mean king problem'' is also naturally formulated in phase space due to the central
role played by mutually unbiased measurements in this context. 

As the Knill--Gottesman theorem states \cite{Gottesman}, the stabilizer sates and their  
evolution under unitary operations of the Clifford group can be efficiently simulated on 
a classical computer. Therefore, one could ask if this class of 
quantum computation also induces a ``classical'' evolution in phase space, i.e. 
if it can be represented by a classical flow mapping phase space points into 
phase space points. For this to
be true, the group of phase space point operators $A(\alpha)$ should be mapped onto itself 
under unitary transformations of the Clifford group. Unfortunately, this is not the 
case. To see this, it is enough to present a counter--example and this is provided by the 
set of Bell states
we studied above. Indeed, Bell states are stabilizer states generated by applying 
operators of the Clifford group to computational states (i.e., a Bell state is obtained
by applying a Hadamard and a controlled--not operator to a computational state). 
As we showed above, while computational states have positive Wigner functions with 
support on the vertical lines, the Wigner function of Bell states have negative 
values for some quantum nets. One could ask if it is possible to choose a quantum 
net such that all stabilizer states have positive Wigner function. The answer 
to this question is, again, negative as the following reasoning shows: The quantum 
net is fully characterized 
by the function $f_\beta={\rm Tr}(T_\beta\ P_{\lambda_\beta})$. On the other hand, a 
stabilizer state represented by a projector $P_S$ is characterized by the function 
$g_\beta={\rm Tr}(P_S\ T_\beta)$. For the Wigner function to be positive $g_\beta$ 
must be equal to $f_\beta$ for all stabilizer states. However, it is simple to show that if
these two functions are identical for a given stabilizer state, one can always construct 
another stabilizer state for which $g_\beta\neq f_\beta$ by using elements of the Clifford group. 
Therefore, for every quantum net one can find some stabilizer states with negative Wigner 
functions. We stress that this does not contradict the conjecture formulated by 
Galv\~{a}o in \cite{Galvao} whose validity would imply that states with negative Wigner
function are necessary for exponential quantum computational speed-up.

\appendix
\section{Representation of the field $GF(2^n)$ by $n\times n$ matrices}

The standard definition of the companion matrix to the binary polynomial
\beq
\pi(x)=r_0 + r_1 x + r_2 x^2 + ...+ r_{n-1} x^{n-1} +x^n
\label {primitive}
\eeq
is 
\beq
M=\left( \begin{array}{cccccc}
           0&1&0&0&\cdots &0 \\
	   0&0&1&0&\cdots &0 \\
	   \vdots& & &\ddots & &\vdots \\ 
	   0& & & & & 1     \\
           r_0&r_1&r_2& \cdots & & r_{n-1}        \end{array}\right).
\eeq
It has the basic property that $\det (M-x)=\pi(x)$ so that $\pi(M)=0$.
Thus $M$ is a root of the polynomial, and if $\pi(x)$ is primitive it can 
be shown \cite{Williams} that  $M^{2^n-1}=1$
and that $M^j\neq 1$ for all $j < 2^n-1$. Thus $M$ can be put in one to one 
correspondence with the abstract elements $\omega^j$ of the field.
 Therefore the set of powers of $M$ gives
a representation of the field in terms of binary matrices with addition and 
multiplication in the field given by ordinary (binary) matrix operations.

The first $n-1$ powers constitute a canonical basis for the field, and
arbitrary elements in this basis can be expanded as
\beq
A=\sum_{i=0}^{n-1} a_i M^i,
\eeq
and are then fully represented by a binary string $ \bfa =(a_0,\ldots,a_{n-1})$.

The action of M  on field elements is then
\beq
  A^\prime=A M= \sum_{i=0}^{n-1} a_i M^{i+1}.
\eeq
Using the fact that $\pi(M)=0$ (equivalent to considering the multiplication
of polynomials modulo $\pi(x)$) we can rewrite this action in matrix form on 
the binary string
\beq
 \bfa^\prime =\bfa M.
 \label{action}
\eeq
The successive powers of $M$ acting on a reference string $\bfa\neq 0$ generates  
a cyclic ordering of the binaries with period $2^n-1$. The opaque looking
definition of the trace of a field element in Eq. (\ref{trace}) becomes here the
ordinary (binary) matrix trace.
 
We are also interested in the action of $M$ on elements expanded in the dual basis.
This is simply given by the transpose $ \tilde{M}$. It should be noted that this matrix also
satisfies $\pi(\tilde{M})=0$ and shares all the primitive properties of $M$.
When acting on a reference binary, it still cycles through all the non-zero binaries,
but in a different order. Table I shows the two different orderings obtained for
$n=2,3$ and 4 qubits for the indicated primitive polynomials.

\begin{table}
$$
 \begin{array}{|rc||rc|}
   \hline
   \multicolumn{2}{|c||}{GF(2^2)} &\multicolumn{2}{c|}{GF(2^4)} \\
   
   \multicolumn{2}{|c||}{\pi(x)=x^2+x+1} &\multicolumn{2}{c|}{\pi(x)=x^4 +x+1}
   \\ & & & \\
   00 & 00       &   0000   & 0000    \\
   10 & 10        &  1000   & 1000    \\
   01 & 01       &   0100   & 0001    \\
   11 & 11       &   0010   & 0010   \\
    &            &   0001   & 0100    \\ 
    \cline{1-2}
   \multicolumn{2}{|c||}{GF(2^3)} & 1100 &  1001  \\
   \multicolumn{2}{|c||}{\pi(x)=x^3+x^2+1} & 0110  & 0011 \\
       &        &  0011   & 0110     \\
   000 &  000   &  1101   & 1101     \\    
   100 &  100   &     1010   & 1010     \\ 
   010 &  001   &    0101  &  0101      \\
   001 &  011   &    1110   & 1011   \\
   101 &  111   &    0111   & 0111   \\ 
   111 &  110   &    1111   & 1111          \\ 
   110 &  101   &    1011   & 1110     \\ 
   011 &   010  &    1001    & 1100      \\         
   \hline  
   \end{array}
$$ 

\caption{The binary orderings obtained in the canonical basis and
its dual for the indicated primitive polynomials. In both columns the reference binary
has been chosen (arbitrarily) as unity.} \label{table1}  
\end{table}

Extensive tables of primitive binary polynomials are available \cite{Stahnke},
and for moderate values of $n$ the above algorithm to generate the sequences
of binaries is easily implemented.
 
We are now interested in obtaining a unitary representation of this action. Consider
the translations $T(\bfa,{\bf 0}) = X^\bfa$ defined in Eq. (\ref{translations}). An
operator $U_\omega$ that achieves this can be constructed
using only two basic gates: $CNOT_{ij}$ and $SWAP_{ij}$. The operator $CNOT_{ij}$ acts 
on qubits $i$ and $j$ transforming the state $|x_i,x_j\rangle$ into $|x_i,x_i+x_j\rangle$. 
The operator $SWAP_{ij}$ interchanges qubits $i$ and $j$. The operator $U_\omega$ 
can then be shown to be \cite{Wootters3}
\begin{equation}
U_\omega=\prod_{j=2}^n CNOT^{r_j}_{1j}\prod_{j=n}^2\ SWAP_{1j}.
\label{uoperator}
\end{equation} 

The operator is completely determined by the coefficients of $\pi(x)$ and its 
action is best understood if we represent it as a circuit acting in the basis
of eigenstates of $Z$, the computational basis.
\begin{figure}  
\includegraphics[width=6cm]{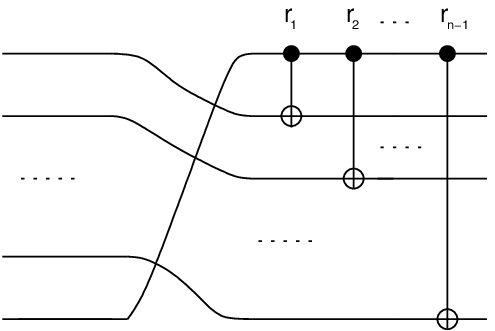}  
\caption{Circuit that implements the action of a primitive element that is a root of
the primitive polynomial $\pi(x)$ in Eq. (\ref{primitive}). The $CNOT$ gates only act if the
corresponding coefficient is non-zero. }  
\label{uoperatorfig}  
\end{figure}

\begin{figure}  
\includegraphics[width=6cm]{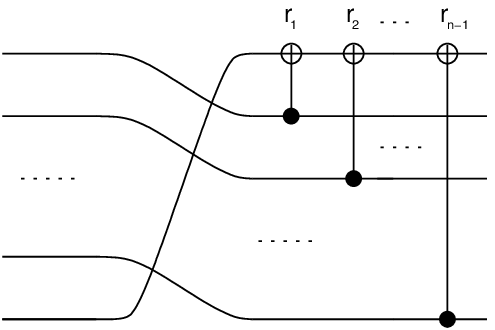}  
\caption{Hadamard transform of the circuit in Fig. \ref{uoperatorfig}. }  
\label{voperator}  
\end{figure}

With the help of some circuit algebra, {\it i.e.} by commuting Pauli matrices with $CNOT$ 
gates we obtain
\beq
U_\omega X^\bfa U_\omega^\dagger = X^{\bfa M},
\label{omegaonx}
\eeq
which is the same as the classical action Eq. (\ref{action}). To 
compute the action of $U_\omega$ on $ Z^{\bfb}$ we apply a tensor product of Hadamard 
transforms
\beq
U_\omega Z^\bfb U_\omega^\dagger= H^{\otimes n} V_\omega X^\bfb {V_\omega}^\dagger H^{\otimes n},
\eeq
where
\beq
V_\omega =   H^{\otimes n}U_\omega H^{\otimes n}.
\eeq
The action of this operator is again best understood as a circuit. The Hadamard gates
just reverse the controls in the $CNOT$'s and the resulting circuit (in the computational
basis) is shown in Fig. \ref{voperator}.

With another bit of circuit algebra we find that 
$ V_\omega X^\bfb V_\omega^\dagger = X^{\bfb \tilde{M}^{-1}}$ and therefore
\beq
U_\omega Z^\bfb U_\omega^\dagger=  Z^{\bfb \tilde{M}^{-1}},
\eeq
The action of $U_\omega$ on the translations is then
\beq
U_\omega T(\bfq,\bfp) U_\omega^\dagger=\pm T(\bfq M, \bfp \tilde{M}^{-1}).
\eeq
The ambiguity in the sign derives from the $\pi/2$ phases and can be calculated but it
needs not concern us here. 
  
In Eq. (\ref{commutingsets}) we defined the ray passing through the point $(\bfa,\bfb)$
as the family of commuting operators $ T(\bfa M^j ,\bfb \tilde{M}^j)$. Acting with
$U_\omega$ on this family we obtain
\beq
U_\omega T(\bfa M^j ,\bfb \tilde{M}^j)U_\omega^\dagger=
               \pm T(\bfa M^{j+1} ,\bfb \tilde{M}^{j-1}). \label{uaction}	       
\eeq

This is now another family corresponding to the ray through 
$(\bfa M, \bfb \tilde{M}^{-1})$. Thus the powers of $U_\omega$ cycle through the $N-1$
``diagonal'' rays, however leaving invariant the horizontal and vertical ones.

\begin{acknowledgments}
JPP and AJR were partially supported by a grant from NSA.
This work was also partially supported with grants from Ubacyt, Anpcyt 03-9000, Conicet and
Fundaci\'on Antorchas. 
\end{acknowledgments}


\begin{thebibliography}{99}

\bibitem{Wigner} E. P. Wigner, Phys. Rev. \textbf{40}, 749 (1932); M. Hillary, R. F. O'Connell, 
M. O. Scully and E. P. Wigner, Phys. Rep. \textbf{106}, 123 (1984).

\bibitem{MPSpra} C. Miquel, J. P. Paz and M. Saraceno, Phys. Rev. A \textbf{65}, 062309 (2002);
P. Bianucci, C. Miquel, J. P. Paz and M. Saraceno, Phys. Lett. A \textbf{297}, 353 (2002). 

\bibitem{Hannay} J. H. Hannay and M. V. Berry, Physica D \textbf{1}, 267 (1980).

\bibitem{Feynman} R. Feynman, ``Negative Probabilities'' in \textit{Quantum Implications: Essays in
Honour of David Bohm}, edited by B. Hiley and D. Peat (Routledge, London, 1987).

\bibitem{Cohen} L. Cohen and M. Scully, Found. Phys. \textbf{16}, 295 (1986).

\bibitem{Wootters1} W. K. Wootters, Ann. Phys. \textbf{176}, 1 (1987).

\bibitem{Galetti} D. Galetti and A. F. R. de Toledo Piza, Physica A \textbf{149}, 267 (1988).

\bibitem{Cohendet} O. Cohendet, P. Combe, M. Sirugue and M. Sirugue-Collin, J. Phys. A \textbf{21},
 2875 (1988).

\bibitem{Leonhardt} U. Leonhardt, Phys. Rev. Lett. \textbf{74}, 4101 (1995); Phys. Rev. A
\textbf{53}, 2998 (1996).

\bibitem{Wootters2} W. K. Wootters, IBM J. Res. Dev. \textbf{48}, 99 (2004); e-print quant-ph/0306135 
(2003).

\bibitem{Wootters3} K. S. Gibbons, M. J. Hoffman and W. K. Wootters, e-print quant-ph/0401155 (2004).

\bibitem{Pazpra} J. P. Paz, Phys. Rev. A \textbf{65}, 062311 (2002).

\bibitem{Lopez} C. C. Lopez and J. P. Paz, Phys. Rev. A \textbf{68}, 052305 (2003).

\bibitem{Bianucci} P. Bianucci, J. P. Paz and M. Saraceno, Phys. Rev. E \textbf{65}, 046226 (2002).

\bibitem{Shepe} B. L\'{e}vi, B. Georgeot and D. L. Shepelyansky, Phys. Rev. E \textbf{67}, 046220 (2003).

\bibitem{MPSnat} C. Miquel, J. P. Paz, M. Saraceno, E. Knill, R. Laflamme and C. Negrevergne,
Nature (London) \textbf{418}, 59 (2002).

\bibitem{PRS03} J. P. Paz, A. J. Roncaglia and M. Saraceno, Phys. Rev. A \textbf{69}, 032312 (2004).

\bibitem{BuzekPra} M. Koniorczyk, V. Bu\v{z}ek and J. Janszky, Phys. Rev. A {\bf 64}, 034301 (2001).

\bibitem{Galvao} E. F. Galv\~{a}o, e-print quant-ph/0405070 (2004).

\bibitem{Gottesman} D. Gottesman, PhD thesis, California Institute of Technology,
e-print quant-ph/9705052 (1997); S. Aaronson and D. Gottesman, e-print quant-ph/0406196 (2004).

\bibitem{Vaidman} L. Vaidman, Y. Aharonov and D. Z. Albert, Phys. Rev. Lett. \textbf{58}, 1385 (1987).

\bibitem{Aharonov} Y. Aharonov and B. -G. Englert, e-print quant-ph/0101065 (2001); B. -G. Englert 
and Y. Aharonov, Phys. Lett. A \textbf{284}, 1 (2001); P. K. Aravind, Z. Naturforsch \textbf{58A}, 
16 (2003); T. Durt, e-print quant-ph/0401037 (2004). 

\bibitem{Ivanovic} I. D. Ivanovi\'c, J. Phys. A \textbf{14}, 3241 (1981).

\bibitem{WFields}W. K. Wootters and B. D. Fields, Ann. Phys. \textbf{191}, 363 (1989).

\bibitem{Bandyo} S. Bandyopadhyay, P. O. Boykin, V. Roychowdhury and F. Vatan,
\textit{Algorithmica} \textbf{34}, 512 (2002); e-print quant-ph/0103162 (2001).

\bibitem{LBZ} J. Lawrence, C. Brunker and A. Zeilinger, Phys. Rev. A \textbf{65}, 032320 (2002);
 S. Chatuvedi, Phys. Rev. A \textbf{65}, 044301 (2002); A. O. Pitterger and M. H. Rubin, e-print
quant-ph/0308142 (2003); T. Durt, e-print quant-ph/0401046 (2004). 

\bibitem{Chuang} \textit{Quantum Information and Computation}, edited by M. Nielsen and I.
Chuang (Cambridge University Press, Cambridge, 2000).

\bibitem{Williams} \textit{Introduction to finite fields and their applications}, edited by R. Lidl 
and H. Niederreiter (Cambridge University Press, Cambridge 1986).

\bibitem{Stahnke} W. Stahnke, Math. Comp. \textbf{27}, 977 (1973).

\end{thebibliography}
\end{document}